\documentclass{article}
\usepackage{graphicx} 
\usepackage{url} 

\usepackage{lineno}
\usepackage{mathtools}
\usepackage{fancyhdr}
\usepackage{latexsym}
\usepackage{amssymb,amsmath,amsfonts} 
\usepackage[mathscr]{eucal}
\usepackage{eqnarray}
\usepackage{graphicx,color}
\usepackage[dvipsnames,svgnames,x11names]{xcolor}
\usepackage{textcomp}
\usepackage{listings}
\usepackage{algorithm2e}
\usepackage{textgreek}
\usepackage{siunitx}
\usepackage{booktabs}
\usepackage{authblk}
\usepackage{appendix}
\usepackage{apacite}

\usepackage[]{changes}

\definechangesauthor[color=Green]{MR}
\definechangesauthor[color=NavyBlue]{HS}
\definechangesauthor[color=Orange]{DK}

\usepackage{todonotes}

\RestyleAlgo{ruled}

\usepackage{geometry}
\providecommand{\keywords}[1]
{
  \small	
  \textbf{\textit{Keywords---}} #1
}

\title{POREMAPS: A finite difference based Porous Media Anisotropic Permeability Solver for Stokes flow}
\author[1,2]{David Krach} 
\author[1]{Matthias Ruf} 
\author[1,2]{Holger Steeb} 
\affil[1]{University of Stuttgart, Institute of Applied Mechanics (CE), Pfaffenwaldring 7, 70569 Stuttgart, Germany.}
\affil[2]{University of Stuttgart, Stuttgart Center for Simulation Science, Pfaffenwaldring 5A, 70569 Stuttgart, Germany.}
\date{\today}

\begin{document}

\urlstyle{same}

\maketitle


\section*{Abstract}
Porous materials are ubiquitous in various engineering and geological applications, where their permeability plays a critical role in viscous fluid flow and transport phenomena. Understanding and characterizing the microscale properties, the effective hydraulic parameters, and also the anisotropy of porous materials are essential for accurate modeling and predicting fluid flow behavior. The study pursues the Digital Rock Physics approach to retrieve intrinsic permeability and its evolution in anisotropic configurations of porous media, which are subjected to pore space alterations. Therefore, we discuss the development and implementation of a computational framework based on the finite difference method to solve the pseudo-unsteady Stokes equations for fluid flow on the pore scale. We present an efficient and highly parallelized implementation of this numerical method for large voxel-based data sets originating from different image-based experimental setups. A comprehensive variety of benchmarks has been conducted to assess and evaluate the performance of the proposed solver. The solver's compatibility with huge domain sizes generated by state-of-the-art imaging techniques is demonstrated. We investigate an open-cell foam undergoing deformation, observing that contrary to initial expectations, no anisotropy emerges. Further, we examine a microfluidic cell experiencing precipitation within its pore space, resulting in clear anisotropic development during the clogging process.

\keywords{Pore-scale resolved modeling, Digital Rock Physics, Permeability, FDM, Anisotropy, X-Ray Comupted Tomography (XRCT)}

\section{Introduction}
Experimental investigations and numerical simulations of fluid flow through porous materials with a large variety of hydro-mechanical properties are of interest in many fields such as hydrosystems modeling, groundwater flow and contaminant transport \cite{bear1987modeling, helmig1997multiphase}, oil reservoir exploitation \cite{aziz1989considerations, cantisano2013relative} and industrial applications like membranes for water desalination \cite{thomas2016computational}. The majority of all investigations operate at the macro, i.e., Darcy scale, and use a coarse-grained continuum theory, which means that individual pore geometries are not physically resolved. Further, the correlation between viscous fluid flow and corresponding driving force is assumed to be linear. Porosity $\phi$ and the intrinsic permeability tensor $ \mathbf{k} = k_{ij}\,\mathbf{e}_i \otimes \mathbf{e}_j$ serve as input parameters for these models. Regimes where the components of the permeability tensor $k_{ij}$ do not linearly depend on the driving force are denoted as non-Darcian and are beyond the scope of this contribution. In general, $\phi$ and effective hydraulic properties like $\mathbf{k}$ of the porous material are highly sensitive to the distinct porous microstructure. This work primarily centers on the assessment and determination of the intrinsic permeability tensor $\mathbf{k}$. There are numerous experimental, semi-analytical and numerical methodologies for permeability determination documented in the literature. Given the expense and complexity of flow experiments and their limitations, and the limited applicability of semi-analytical models it is preferable to assess these parameters with computer simulations. Advances in three-dimensional (3D) imaging techniques like micro X-Ray Computed To\-mo\-gra\-phy ({\textmu}XRCT) and the meanwhile widespread availability of appropriate devices \cite{Withers2021, wildenschild2002using, wildenschild2013x, ruf2020open} allow to fully resolve $3$D pore geometries at high-resolution and allow for the flow problem to be solved directly at the pore scale for many porous materials \cite{blunt2013pore}. Moreover, two-dimensional (2D) data, e.g., generated from images of microfluidic experiments acquired by optical microscopy, extended in the third spatial direction can provide further input data \cite{hommel2022effects, weinhardt2022spatiotemporal}.

\noindent Deformation of the pore space or precipitation and a corresponding decrease in porosity have a direct influence on hydro-mechanical properties of a material. Certainly, the process of experimentally determining anisotropic permeability in the course of an experiment is inherently intricate. Acquiring the complete permeability tensor $\textbf{k}$ is mostly impossible. Nonetheless, this investigation is particularly significant due to the non-trivial implications of deformation or precipitation on the principal directions of hydraulic properties or the material's anisotropic characteristics. \noindent In the literature, numerical solvers are applied to porous materials undergoing changes in pore space, cf. for geochemical alteration like calcite precipitation \cite{noiriel2016effects}, carbonate precipitation for carbon capture and storage applications \cite{jiang2014changes} or external strain causing deformation \cite{bakhshian2016computer,hilliard2024modeling}. Another field of interest is the benchmark of specially designed materials whereby the permeability can be controlled via the manufacturing parameters on Al-Cu alloys \cite{bernard2005permeability}. 
Due to the limitations of experimentally viable boundary conditions, e.g. undrained boundary conditions along the cylinder of the sample in conventional triaxial cells, a priori assumption has to be made for the orientation of principal directions of $\mathbf{k}$.
The utilization of the aforementioned imaging techniques combined with computer simulations provides a way for achieving this in a systematic way while circumventing any a priori assumption. However, in order to determine changes in anisotropy or computing porosity-permeability relationships, for example, many data points and correspondingly many individual simulations are required. 
While various numerical methods like Finite Element Method (FEM) \cite{borujeni2013effects, chareyre2012pore}, Finite Volume Method (FVM) \cite{chareyre2012pore,Liao2024} and Lattice-Boltzmann Methods (LBM) \cite{borujeni2013effects, liu2014characterisation}, Smoothed Particle Hydrodynamics (SPH)\cite{osorno2021cross, tartakovsky2016smoothed, holmes2016characterizing} or Pore Network Modeling (PNM) \cite{dong2009pore, piovesan2019pore, blunt2017multiphase} have been employed to simulate pore-scale fluid flow, each method comes with its own set of limitations and advantages.  In this work, we present an efficient and reliable solver for the Stokes equations by implementing a Finite Difference Method (FDM) based algorithm.
In terms of numerical efficiency, the solver is tuned for voxel-based cartesian grids as directly obtained from image-based characterization methods like XRCT or microfluidics.
It stands out as a versatile and well-established choice due to its straightforward implementation and suitability for complex geometries \cite{Peyret1983, Bentz2007, Gerke2018, lemaitre1990fractal, adler1990flow}. In addition, FDM is a simple approach in terms of the algorithm, has advantages when using regular grids, and can be effectively parallelized due to its local nature.

\noindent The objective is to develop a resilient and modular open-source tool that advances our comprehension of fluid flow at the pore-scale and addresses the issue of evolving anisotropies in porous media under varying conditions. The software tool should be able to handle state-of-the-art data sets from XRCT scans (up to 2000$^3$ voxels) on various computing architectures and should allow for further extensions. 
Besides research codes which do not allow for further extensions (only executable of the software available \cite{Gerke2018}) or commercial solvers such as GeoDict \cite{geodict2023}, there are common open-source packages such as OpenFoam, e.g. used in \citeA{icardi2014pore, guibert2016comparison}, which are extremely flexible but therefore not tuned for the mentioned demands. Others, like the tool from the National Institute for Standards and Technology \cite{Bentz2007} which is written in Fortran, are not multi-node parallelized.  
The rationale behind developing a new solver, despite the existence of current solutions, is multilayered. We present a FDM solver, exclusively dependent on the Message Passing Interface (MPI), that is fully open source and platform independent and works directly on binarized data sets which essentially is an important tool in porous media research. Moreover, its full integration into existing experimental setups streamlines the research process, promoting a more cohesive and comprehensive approach to investigate complex phenomena. Since large domains (state-of-the-art XRCT data-sets comprise up to $10^9$--$10^{10}$ voxels) have to be simulated, in combination with several snapshots for time-resolved investigations, the efficiency and performance as well as the simplicity of the presented FDM solver is the main focus. The code is written in procedural, functional C++ and is parallelized with the MPI to run on multi-node, multi-core CPU systems. 

\noindent For these reasons, in addition to the permeability determination derivation (Sec.~\ref{sec:K}), the mathematical basics (Sec.~\ref{sec:math_bg}), and the numerical principles (Sec.~\ref{sec:FD}), it is particularly important for us to address the implementation and technical aspects (Sec.~\ref{sec:technical}) as well as to provide a detailed validation against various benchmark cases (Sec.~\ref{sec:benchmarks}). In Sec.~\ref{sec:applications}, we demonstrate the developed solver to investigate two distinct materials characterized by alterations in their pore space. These alterations are anticipated to result in changes not only in the magnitude of the permeability but also in the permeability anisotropy ratio. Specifically, our study focuses on an open-cell foam, which defies expectations by not exhibiting anisotropy as it undergoes deformation. Additionally, we investigate a porous micro-structure exposed to mineral precipitation and subsequently clogging, revealing a development of anisotropy that is notably influenced by the boundary conditions of the underlying experiment.

\section{Permeability tensor and principal permeabilities}\label{sec:K}

\noindent Permeability is defined as a proportionality factor between pressure gradient across the examined sample and fluid fluxes ($\mathrm{grad}\, p \propto \mathbf{q}$). For homogeneous materials, that are known to have isotropic material behavior a scalar value is used for the hydraulic permeability. Intrinsic permeability quantifies viscous losses in continuum-based Darcy-type models. In the generic case, the effective permeability $\mathbf{k}$ is a second order tensor

\begin{equation}\label{eq:3dK}
    \mathbf{k} = k_{ij} \, \mathbf{e}_i \otimes \mathbf{e}_j =  \begin{bmatrix}
        k_{11} & k_{12} & k_{13}\\
        k_{21} &  k_{22} & k_{23} \\
        k_{31} & k_{32} & k_{33}
\end{bmatrix} \mathbf{e}_i \otimes \mathbf{e}_j
\end{equation} 

\noindent where $k_{ij}$ are the components and the cartesian basis vectors $\mathbf{e}_i$ build the tensorial basis through its dyadic product. Bold notation is employed to represent tensors and vectors, with the implicit assumption that equations in index notation abide by Einstein's summation convention. The coefficient matrix is symmetric and positive definite \cite{bear1988dynamics, scheidegger1957physics}. In contrast to the experimental determination of permeability, the numerical approach is capable of computing secondary diagonal elements of $k_{ij}$, which is required for a complete hydraulic characterization.
Solving the characteristic polynomial $\mathrm{det} \,(\mathbf{k} - \lambda \mathbf{I}) \, = 0$ for the eigenvalues $\lambda_i$ (also referred to principal permeabilities $k_{I}$, $k_{II}$, $k_{III}$) we obtain the permeability tensor in its spectral form

\begin{equation}\label{eq:3dK_eigen}
    \mathbf{k} =  \sum_{i = 1}^3 \lambda_i \, \tilde{\mathbf{e}}_i \otimes \tilde{\mathbf{e}}_i \, ,
\end{equation}

\noindent expressed in a basis system with the principle axes $\tilde{\mathbf{e}}_i$, that are computed by solving $(\mathbf{k} - \lambda_i \mathbf{I}) \cdot \tilde{\mathbf{e}}_i = \mathbf{0}$. $\mathbf{I}$ is the second order identity tensor. The orientation of the principal axes with respect to the basis system $\mathbf{e}_i$ is computed via the rotation tensor $\mathbf{R} = R_{ik} \, (\mathbf{e}_i \otimes \mathbf{e}_k)$ whereby $\tilde{\mathbf{e}}_i = \mathbf{R} \cdot \mathbf{e}_i$ and $R_{ki} = \mathrm{cos} \, \sphericalangle \, (\mathbf{e}_k, \tilde{\mathbf{e}}_i)$ hold. Determining the principal permeabilities enables categorization of the material into one of the following classes:
\begin{enumerate}
    \item \textbf{Isotropy}: $k_{I} = k_{II} = k_{III}$\\
            Same hydraulic properties in all three principal directions.
    \item \textbf{Orthotropy}: $k_{I} \neq k_{II} \quad \wedge \quad k_{I} \neq k_{III} \quad \wedge \quad k_{II} \neq k_{III}$ \\
    Unique and independent hydraulic properties in three mutually perpendicular directions.
    \item \textbf{Transverse anisotropy}: $k_{I} \neq k_{II} = k_{III} \quad \vee \quad k_{I} = k_{II} \neq k_{III} \quad \vee \quad  k_{I} = k_{III} \neq k_{II}$\\
    Same hydraulic properties in one plane. Thus, there are two independent constants in the permeability tensor. Typical examples of this material class are wood, unidirectional fiber composites or sedimentary sandstones. 
\end{enumerate}

\section{Governing equations on the pore scale}\label{sec:math_bg}
Pore scale refers to the length scale at which individual pores and their geometrical features, such as pore size and shape, are significant. At this scale, the fluid flow pattern is strongly influenced by the morphology of the pores, as well as the viscous momentum interactions between the fluid phase and the solid skeleton. Pore scale simulations are used to study fundamental aspects of porous media flow as well as the determination of properties, used in coarse-grained Darcy scale methods. We are interested on solving for intrinsic effective permeability and therefore consider solving fluid flow at stationary creeping flow conditions. Accordingly, the following applies for the Reynolds number  
\begin{equation}\label{eq:reynoldsnumber}
    \mathrm{Re} \, = \, \frac{\rho_0 \, \mathscr{V} \, \mathscr{L}}{\mu} \,  < 1  \qquad [-] \, , 
\end{equation}
\noindent and the fundamental equations to be solved are the Stokes equations consisting of the balance of linear momentum for an incompressible Newtonian fluid  \cite{Peyret1983} 
\begin{equation}\label{eq:stokes_equtaions_physical}
    \mathbf{0} = \mu \, \mathrm{div}(\mathrm{grad} \, \mathbf{v}) - \mathrm{grad}\, p \, 
\end{equation}
and the balance of mass 
\begin{equation}\label{eq:mass_balance}
    \mathrm{div} \,\mathbf{v} = 0 
\end{equation}
where $\rho$, $\rho_0$, $\mathbf{v}$, $\mu$, $p$, $\mathscr{V}$, $\mathscr{L}$ are the fluid density, the rest density of the fluid, fluid velocity vector, (constant) dynamic viscosity of the fluid, pore fluid pressure and the characteristic velocity and length, respectively. No body force is used. In porous media $\mathscr{L}$ is assumed as the characteristic (mean) pore diameter and $\mathscr{V}$ is calculated from the applied/measured flux $\mathbf{q}$. In addition to characteristic properties, indicated by notation in script typestyle, to scale physical quantities we denote dimensionless variables and operators by $(\bullet)^\ast$. It is well known that solving Eqs.~(\ref{eq:stokes_equtaions_physical}) and (\ref{eq:mass_balance}) causes problems due to the special role of pressure. Therefore, this set of governing equations is extended with an artificial time derivative and a so-called pseudo-unsteady method is used \cite{Peyret1983}. In this manner, for the numerical implementation, we do not fulfill the incompressibility condition as an algebraic constraint. Instead a artificial compressibility formulation will considered with an equation of state for the fluid pressure $p(\rho)$. Hence, we obtain for the nondimensional equations to be solved
\begin{equation}
    \rho^\ast \overset{\ast}{(\mathbf{v}^\ast )}\!^{\boldsymbol{\cdot}} = \frac{1}{\mathrm{Re}} \, \overset{\ast}{\mathrm{div}} \, \overset{\ast}{\mathrm{grad}} \, (\mathbf{v}^\ast) - \overset{\ast}{\mathrm{grad}} (p^\ast) \quad \text{and} \quad  \overset{\ast}{(p^\ast)}\!^{\boldsymbol{\cdot}} = - {c^\ast}^2 \overset{\ast}{\mathrm{div}} (\mathbf{v}^\ast) \, ,
\end{equation}
\noindent where $c^\ast$ is the dimensionless speed of sound also referred to artificial compressibility parameter. This numerical approach solves the steady-state Stokes equations by transforming them into a pseudo-time dependent problem which is feasible when dealing with low Reynolds number fluid flows in porous media. It provides a computationally efficient alternative to directly solving the steady-state equations and it can be shown that the solution converges for $t \rightarrow \infty $ to the steady-state solution of the original problem \cite{Chorin1967, Gerke2018, Peyret1983}. The detailed nondimesionalization with introduction of all quantities, dimensionless differential operators and constitutive equations can be found in Appendix~\ref{appendix:A}. 

\section{Finite difference scheme for Stokes equations}\label{sec:FD}

\paragraph{Simulation parametrization:}
We introduce the reference length $\mathscr{L} = 1\,  \mathrm{vx}$ (voxel) and the reference velocity $\mathscr{V} = \frac{1 \mathrm{vx}}{\mathrm{s}}$. As driving force a constant pressure gradient across the domain is employed and as initial conditions we set $\overset{\ast}{\mathrm{grad}} (p^\ast) = \frac{1}{\mathrm{vx}}$. Furthermore, no body force is present ($\mathbf{b} = \mathbf{0}$). We require small Reynolds number and fix $\mathrm{Re} = 0.01$ and ${c^\ast}^2 = 1.5 \times 10^6$ as in \citeA{Bentz2007}. Note that the dimension ${c^\ast}^2\approx 10^6$ fits well to the physical ratio of the speed of sound in water to the assumed characteristic velocity of creeping flow conditions.

\paragraph{Grid:}
Space is discretized with a regular staggered Marker-and-Cell (MAC) grid \cite{harlow1965numerical}, where fluid velocities are stored on the faces of the cells, while the pressure values are stored at the cell centers. Storing different quantities on different locations within each cell, allowing for an efficient and accurate computation of the pressure gradients and additionally for the exact modeling of no-penetration conditions at the fluid-solid interfaces. 

\begin{figure}[h]
\centering
\includegraphics[width=0.90\textwidth]{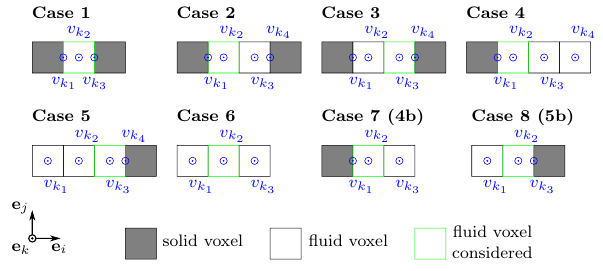}
\caption{Different cases for the identification of the voxel neighborhood to calculate the second-order derivatives ($\frac{\partial^2 v_k}{\partial x_i^2}$) in the perpendicular directions to the considered flow velocity direction for $i,j,k =  (1,2,3), (2,3,1), (3,1,2)$.}
\label{fig:neighborhoodCases}
\end{figure}

\paragraph{Boundary in the numerically considered domain and second order derivatives:}
For the outer boundaries we apply periodic conditions to simulate a periodic or repeating behavior of the physical system in a unit cell being modeled or use no-slip no-penetration boundary conditions initially assumed to be not suitable for  anisotropy investigations since the transverse flow would be zero. The outer boundary in direction of the pressure gradient is always periodic for the local flux. In main flow direction we employ fixed pressure conditions. Fluid-solid boundaries correspond to the pixel boundaries of the binary 3D image and we use no-slip no-penetration boundary conditions on the interfaces, cf. Fig.~\ref{fig:neighborhoodCases}. Second-order derivatives perpendicular to the local flow direction are analytically determined based on Taylor series approximations. This ensures that the no-slip condition is fulfilled on the fluid-solid voxel surfaces. We distinguish 8 basis cases, whereas the last two ones are special cases of cases 4 and 5, and are only required for the second voxel layers used for the boundary conditions. Therefore, we have theoretically the same number of 6 cases as in \citeA{Bentz2007,Gerke2018}. In Fig.~\ref{fig:neighborhoodCases} the eight distinguished cases are exemplary illustrated for the flow direction $\mathbf{e}_k$ and the perpendicular direction $\mathbf{e}_i$. 

\noindent Based on the velocity definitions in Fig.~\ref{fig:neighborhoodCases}, the second-order derivatives for the different cases can be computed by solving a resulting system of equations. For instance, for case 2 we get a set of three equations to compute $\frac{\partial^2 v_k}{\partial x_i^2}$ as follows:
\begin{align}
	\left.\begin{aligned}
	v_{k_1} &\approx v_{k_2} - \frac{1}{2} \frac{\partial v_k}{\partial x_i} + \frac{1}{8} \frac{\partial^2 v_k}{\partial x_i^2} - \frac{1}{48} \frac{\partial^3 v_k}{\partial x_i^3} = 0 \\
	v_{k_3} &\approx v_{k_2} + \frac{\partial v_k}{\partial x_i} + \frac{1}{2} \frac{\partial^2 v_k}{\partial x_i^2} + \frac{1}{6} \frac{\partial^3 v_k}{\partial x_i^3} \\
	v_{k_4} &\approx v_{k_2} + \frac{3}{2} \frac{\partial v_k}{\partial x_i} + \frac{9}{8} \frac{\partial^2 v_k}{\partial x_i^2} + \frac{9}{16} \frac{\partial^3 v_k}{\partial x_i^3} = 0
	\end{aligned}
	\right\}
\quad \leadsto \frac{\partial^2 v_k}{\partial x_i^2} = \frac{8}{3} v_{k_3} - \frac{16}{3} v_{k_2}
\end{align}
An overview of all cases of neighborhood consideration can be found in the Appendix~\ref{appendix:B}.

\begin{figure}[h!]
    \centering
    \includegraphics[width=0.8\textwidth]{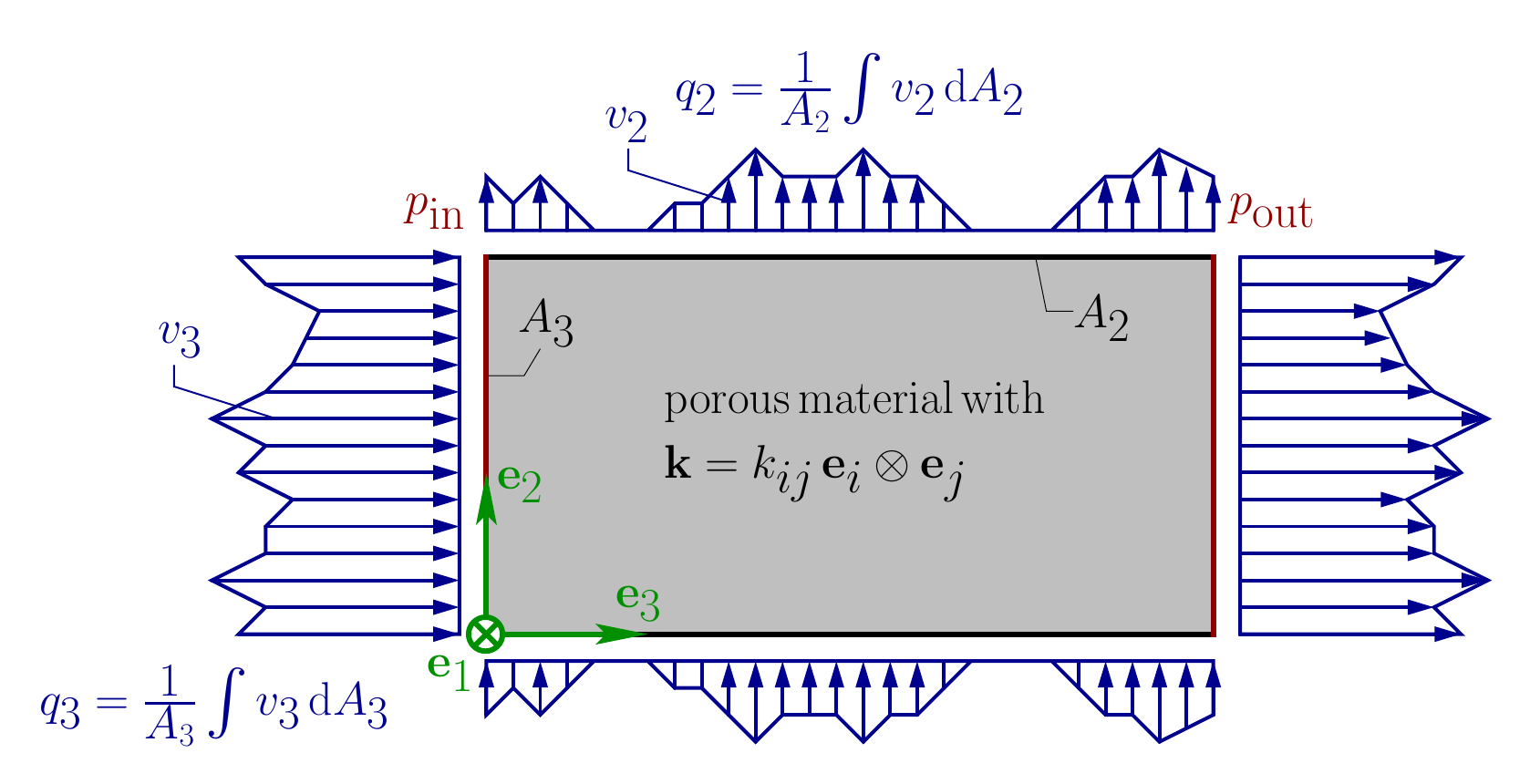}
    \caption{Illustration of the components of Darcy's law with the given boundary values for the pressure $p_\mathrm{in}$ and $p_\mathrm{out}$ used to compute the gradient $h_{3}$. Volumetric fluxes $q_2$, $q_3$ are evaluated over the respective boundaries $A_2$, $A_3$. In $\mathbf{e}_1$-direction, the volumetric flux is not shown here, but is handled analogously. With given pressure gradients $h_{3} \neq 0$ and $h_{1} = h_{2} = 0$, this represents case \textit{c)} in Eq.~(\ref{eq:cases_pressure}).}
    \label{fig:perm_computation}
\end{figure}

\paragraph{Permeability computation:}
In the so-called creeping flow regime, characterized by low Reynolds numbers, $\mathrm{Re} < 1.0$, it is valid to employ Darcy's law \cite{darcy1856fontaines} 
\begin{equation}\label{eq:darcy}
    \mathbf{q} = \frac{1}{\mu} \mathbf{k} \cdot \mathbf{h} \quad \rightarrow \quad  q_i \mathbf{e}_i = \frac{1}{\mu} k_{ij} (\mathbf{e}_i \otimes \mathbf{e}_j) \cdot h_k \mathbf{e}_k = \frac{1}{\mu} k_{ik} h_k \mathbf{e}_i 
\end{equation}
to compute the entries of the second order permeability tensor where $h_i = - \frac{\partial p}{\partial x_i} = - p_{,i}$ is the pressure gradient. In order to determine the nine entries in the coefficient matrix of the permeability tensor, three numerical simulations must be performed. The following pressure gradients $h_i$ are specified for the different simulations 
\begin{equation}\label{eq:cases_pressure}
    a) \,\, h_{1} \neq 0\, \wedge \, h_{2} = h_{3} = 0\, ; \quad 
    b) \,\, h_{2} \neq 0\, \wedge \, h_{1} = h_{3} = 0\, ; \quad
    c) \,\, h_{3} \neq 0\, \wedge \, h_{1} = h_{2} = 0\, ; 
\end{equation}
and we measure three fluxes $q_i$ for each case. For case \textit{c)}, all properties are illustrated in Fig.~\ref{fig:perm_computation}. By comparing the coefficients (Eq.~(\ref{eq:darcy}), right), nine equations are obtained for nine entries of the coefficient matrix of the permeability tensor. A detailed list of the equations for determining the complete permeability tensor $\mathbf{k}$ can be found in Appendix~\ref{appendix:C}.

\section{Computational aspects}\label{sec:technical}

\subsection{Processing input data}
Binarized 3D image data (8-bit file format) distinguishing between solid phase and pore space are employed as input for the solver. Typically performed image pre-processing steps depending on the exact imaging technique  (e.g, denoising and segmentation) are well-documented in the literature, e.g., \citeA{andra2013digital, andra2013digitalb, Burger2016, Iassonov2009, Russ2016, Schlueter2014, Tuller2013} and will not be further elaborated upon here. Depending on the boundary conditions and material, a few pre-processing steps need to be conducted, such as mirroring for symmetric periodicity, domain cropping, or the elimination of disconnected pore spaces. This depends upon the specific problem at hand, a topic that will be delved into within the dedicated application sections. The solver operates under the assumption that the percolation condition is met.

\subsection{Implementation} 

The solver is completely implemented in C++ and parallelized with OpenMPI (version $4.1.5$) \cite{mpi40} to employ it on distributed memory architectures. Special emphasis is put on keeping the code as simple as possible. For the domain decomposition we use a communicator on a Cartesian topology (\texttt{MPI\textunderscore Cart\textunderscore create}) which is particularly well suited for regular meshes and 3D geometries. The file IO is fully parallelized and communication between the ranks is executed by blocking send-receive operations. We work directly with binary data which is RAM-efficient allows for a concise. The source code of POREMAPS is published in \citeA{krach2024a}.

\SetKwComment{Comment}{/* }{ */}

\begin{algorithm}
\caption{Program Flow Stokes Solver}\label{alg:prog_flow}
\KwData{Binarized 3D image data}
\KwResult{Velocity and pressure fields}
Initalize MPI;\\
Domain decomposition, read partial domains, add halos; \\
Impose initial pressure gradient;\\
Evaluate neighborhood; \\

\While{$\mathrm{div} \, \mathbf{v} > \varepsilon$}{
  Compute velocity field $\mathbf{v}_{i,j,k}$;\\
  Compute pressure field $p_{i,j,k}$;\\
 
  \If{$i \mod CommFrequency \equiv 0$ }{
     Communicate halos 
  }
  Evaluate permeability and convergence criterion $\varepsilon$;\\
  Write log file;
}
Write pressure and velocity fields;
\end{algorithm}

The inherent domain decomposition of MPI optimizes communication by embedding the virtual topology onto the physical machine as efficient as possible but is not necessarily suitable for domains with high aspect ratios (see Sec.~\ref{ssec:dev_anios_precipitation}). Therefore, it is possible to include the desired number of ranks in each direction directly in the input file.

\subsection{Scaling}
The scalability of FDM codes has been studied extensively in the literature. Several factors contribute to weak scalability, including the communication overhead, memory requirements, and load balancing. For the scaling test we use a $50^3$ voxel regular sphere packing per core and the hardware of the experimental compute cluster "ehlers" of the EXC 2075 "SimTech" Cluster of Excellence (University of Stuttgart). The CPU partition is comprised of $8$ nodes with $128$ cores ($2 \times 64$ cores, AMD EPYC 7702) each and $200\,\mathrm{Gb}/\mathrm{s}$ Infiniband interconnect and $2 \, \mathrm{TB}$ of RAM.

\begin{figure}[h]
    \centering
    \includegraphics[width=0.49\textwidth]{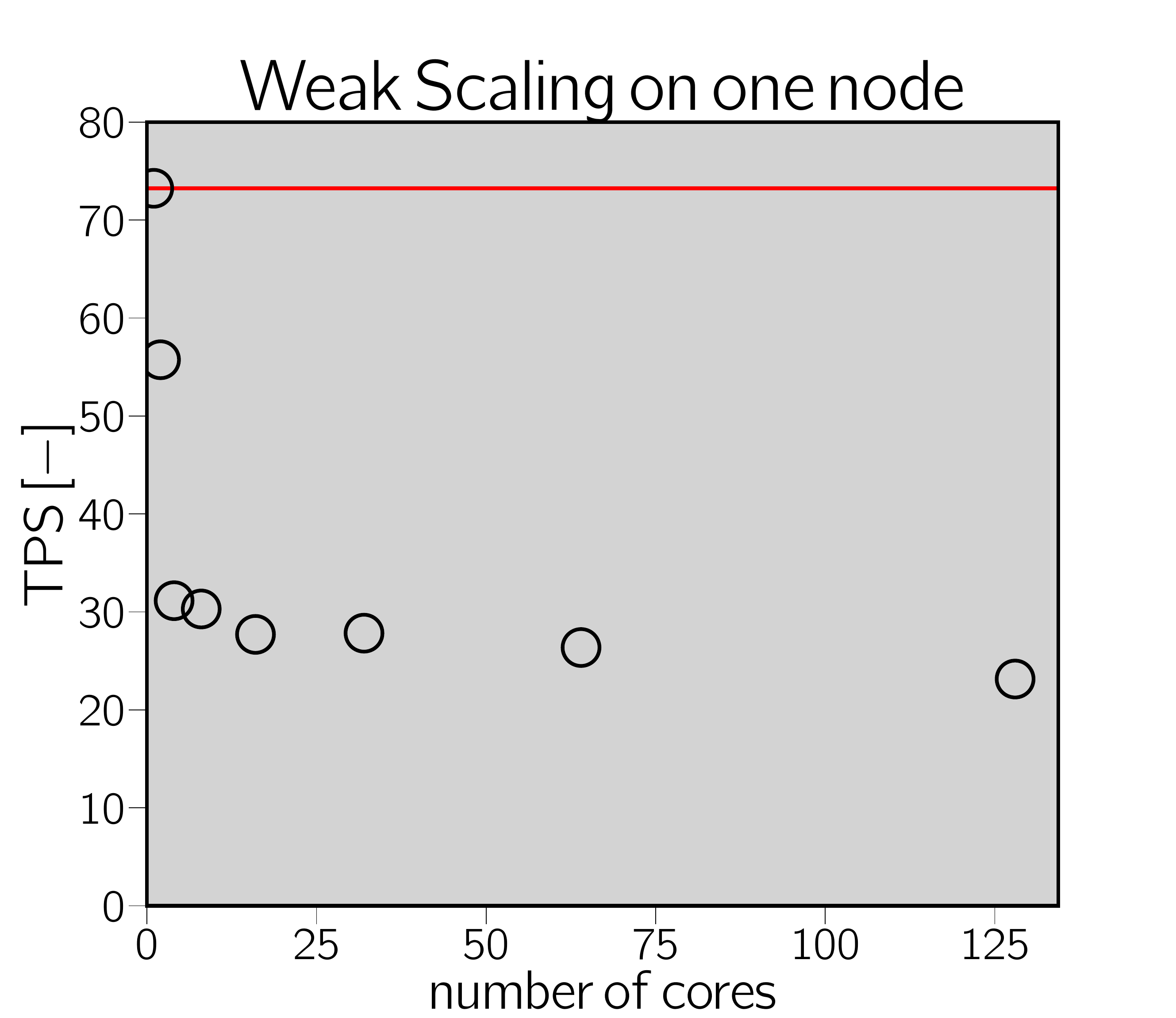}
    \includegraphics[width=0.49\textwidth]{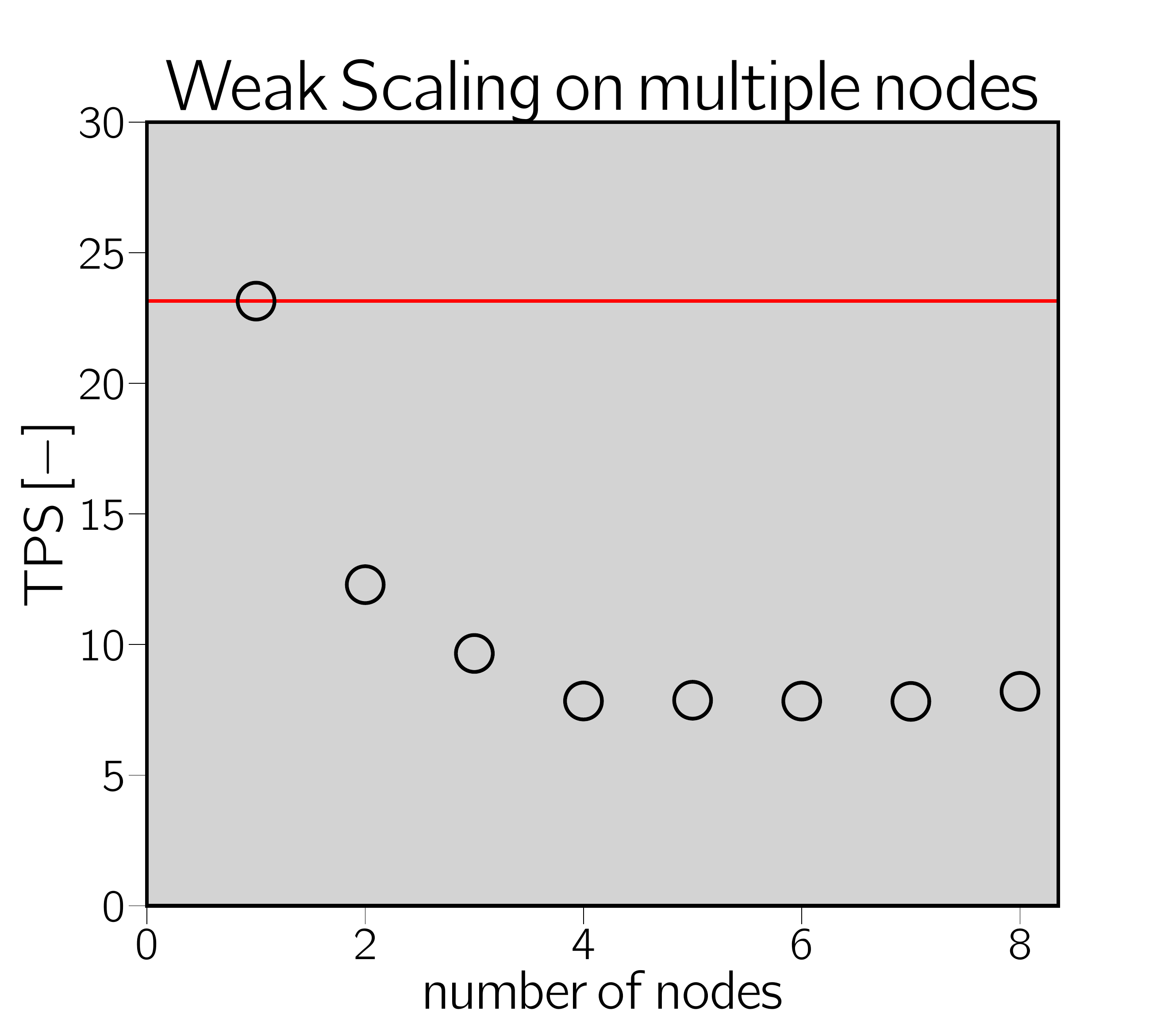}
    \caption{Weak scaling on one node using 1, 2, 4, 8, 16, 32, 64, 128 cores (left) and 1--8 nodes with 128 cores each (right).}
    \label{fig:weak_scaling}
\end{figure}
\noindent The algorithm is very communication-heavy. The influence of increasing core-to-core (Fig.~\ref{fig:weak_scaling}, left) or node-to-node communication (Fig.~\ref{fig:weak_scaling}, right) causes a decrease in performance. However, as soon as communication in all three spatial directions is required ($> 8$ cores), the computed time steps per second (TPS) are almost constant. Accordingly, the results are satisfactory for the problems at hand. Considering targeted domain sizes, tests beyond 1024 cores are not significant for our applications and therefore not considered here. 

\section{Benchmarks}\label{sec:benchmarks}

The code has been developed to be able to study complex, heterogeneous porous materials. To justify the application of the solver to diverse domains, a multi-layered benchmark and validation program is conducted. We validate the code against analytical solutions such as Poiseuille flow and channel flow as well as empirical relations such as the Kozeny-Carman equation for sphere packings with different porosities. We further evaluate our workflow for the determination of the anisotropy and the secondary diagonal elements of the permeability tensor. Finally, we compare our code with other solvers like LBM, FEM or mathematical homogenization for different types of porous materials.

\subsection{Hagen-Poiseuille equation}\label{ssec:hagenpoi}
The laminar creeping flow of a Newtonian fluid through a pipe is a standard benchmark for CFD codes. The tube is per definition periodic in direction of the pressure gradient. We compare results for different resolutions and the analytical Hagen-Poiseuille equation presented in \citeA{batchelor1967introduction}
\begin{equation}
    \label{eq:hagen_poi_analytic}
    v_3(r) = \frac{\Delta p}{4L\mu} (R^2 -r^2) 
\end{equation}
with the radial coordinate $r$. Length and radius of the tube are $L = \SI{0.01}{\meter}$ and $R = \SI{0.001}{\meter} $. The no-slip condition on the fluid-solid interface results in $ v_3(r = R) = 0$. 

\begin{figure}[h]
    \centering
    \includegraphics[width=0.95\textwidth]{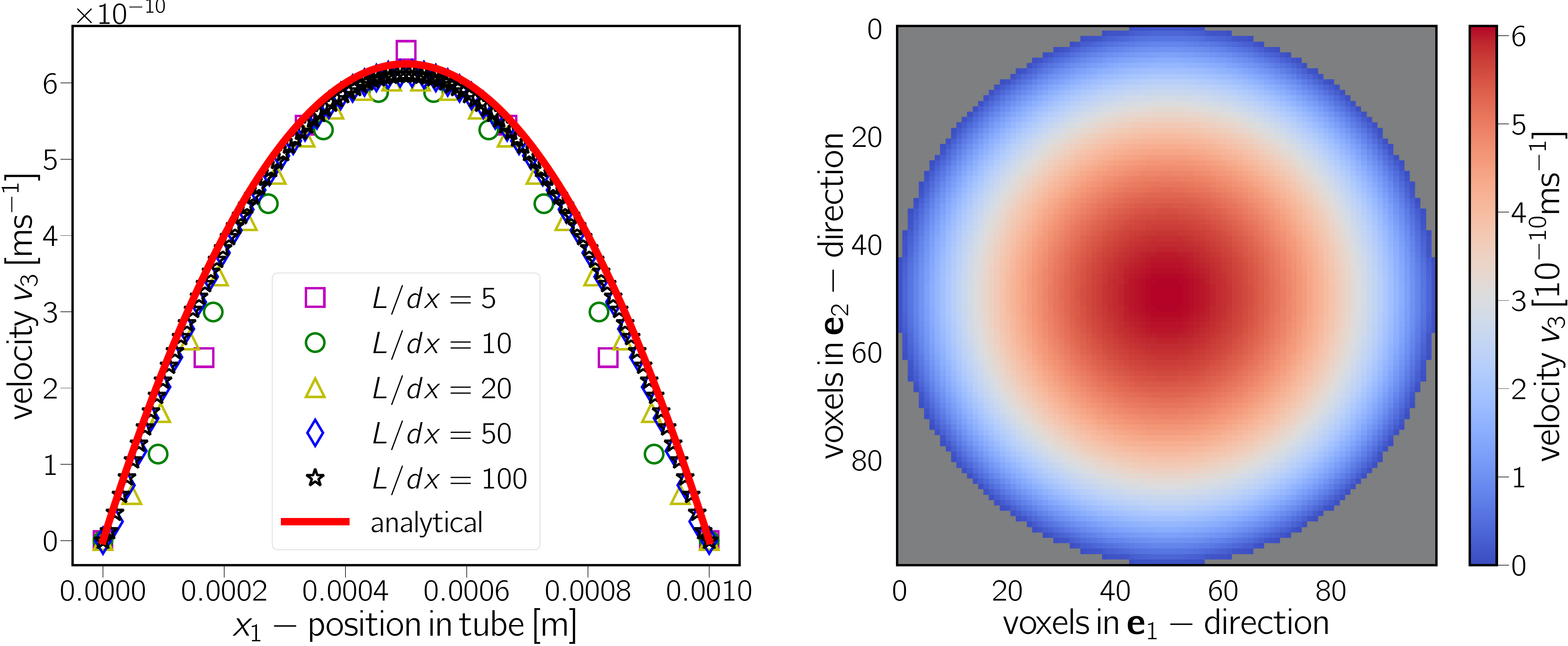}
    \caption{Resolution dependent velocity profiles through a tube compared to the analytical solution (left) and velocity pattern in one cross-section (right).}
    \label{fig:tube_benchmark}
\end{figure}
We compare simulation results obtained for different resolutions ($L/\mathrm{d}x = \{ 5, 10, 20, 50, 100 \}$) and present the velocity profiles through the center of the tube, see Fig.~\ref{fig:tube_benchmark}, left. All resolutions deliver very similar solutions and are in line with the analytical solution Eq.~(\ref{eq:hagen_poi_analytic}). Fig.~\ref{fig:tube_benchmark}, right, gives the cross-section through the tube visualizing the radially symmetric velocity pattern.

\subsection{Channel flow -- rectangular cross section}
To benchmark slightly more complex structures, we analyze the flow through a rectangular channel. It has an advantage over the tube in Sec.~\ref{ssec:hagenpoi}, since it can be discretized on a cubic lattice without introducing a discretization error. The benchmark provides information about the resolution at which we can expect valid results from the solver. That is of particular interest for the study of microfluidic experiments in order to resolve the channels sufficiently.
\begin{figure}[h!]
    \centering
    \includegraphics[width=0.95\textwidth]{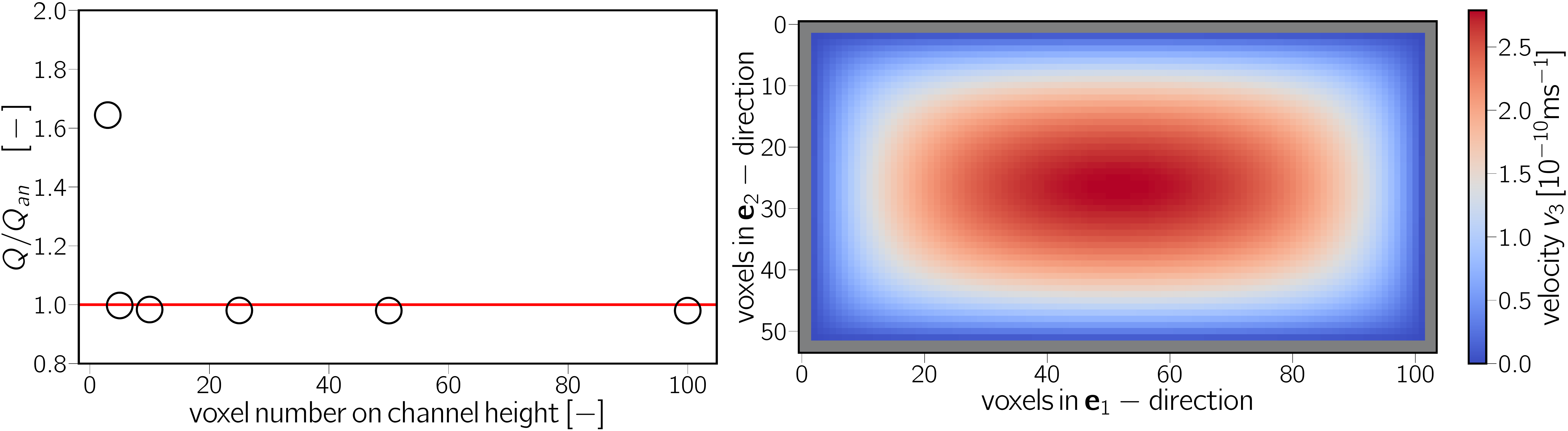}
    \caption{Deviation of the volumetric flux $Q$ from its analytical solution $Q_{an}$ according to \protect\citeA{white2006viscous} through rectangular cross section channel for different resolutions (left) and corresponding velocity pattern (right).} 
    \label{fig:rect_channel_benchmark}
\end{figure}

\noindent Presented are simulation results for different resolutions ($h/\mathrm{d}x = \{ 3, 5, 10, 25, 50, 100 \}$) for the channel height $h$. The width of the channel equals $b = 2h$. Already with a resolution of the channel cross-section using $5 \times 10$\,voxels, the volumetric flow rate $Q = q \cdot A$ corresponds with results for higher resolutions. Although the solution depends on the discretization, it converges already for low resolutions.

\subsection{Regular sphere packings}
Due to the periodicity of regular sphere packings, the simulation of flow can be reduced to a cubic unit cell of side-length $L$. It is therefore possible to simulate representative porous structures without pushing the domain size too far. In addition, the permeability in direction of the cubic unit cell edges is identical and semi-analytical estimates exist. Therefore, we investigate differently arranged sphere packings (Simple Cubic (SC), Body-Centered Cubic (BCC), Face-Centered Cubic (FCC)) for a sweep over a wide range for the porosity $\phi = \{ 0.1, 0.2, 0.3, 0.4, 0.5, 0.6, 0.7, 0.8, 0.9 \}$. In addition, all simulations are run at three different resolutions $L/\mathrm{d}x = \{ 60, 100, 200 \}$. The results are compared with permeability estimates $k_1^{KC}$ by the semi-analytical, semi-empirical Kozeny-Carman equation \cite{carman1997fluid, kozeny1927uber}

\begin{equation}\label{eq:kozeny-carman}
k_1^{KC} = \frac{D^2}{c_{KC}} \frac{\phi^3}{(1-\phi)^2}\, ,
\end{equation}

\noindent where $D$, $c_{KC}$ are the sphere diameter and the Kozeny-Carman constant for which the value $c_{KC} = 180$ is set. In addition, the Stokes solution is further compared with results from a SPH solver that performs simulations based on weakly compressible Navier-Stokes equations \cite{osorno2021cross}. 

\begin{figure}[h!]
    \centering
    \includegraphics[width=0.99\textwidth]{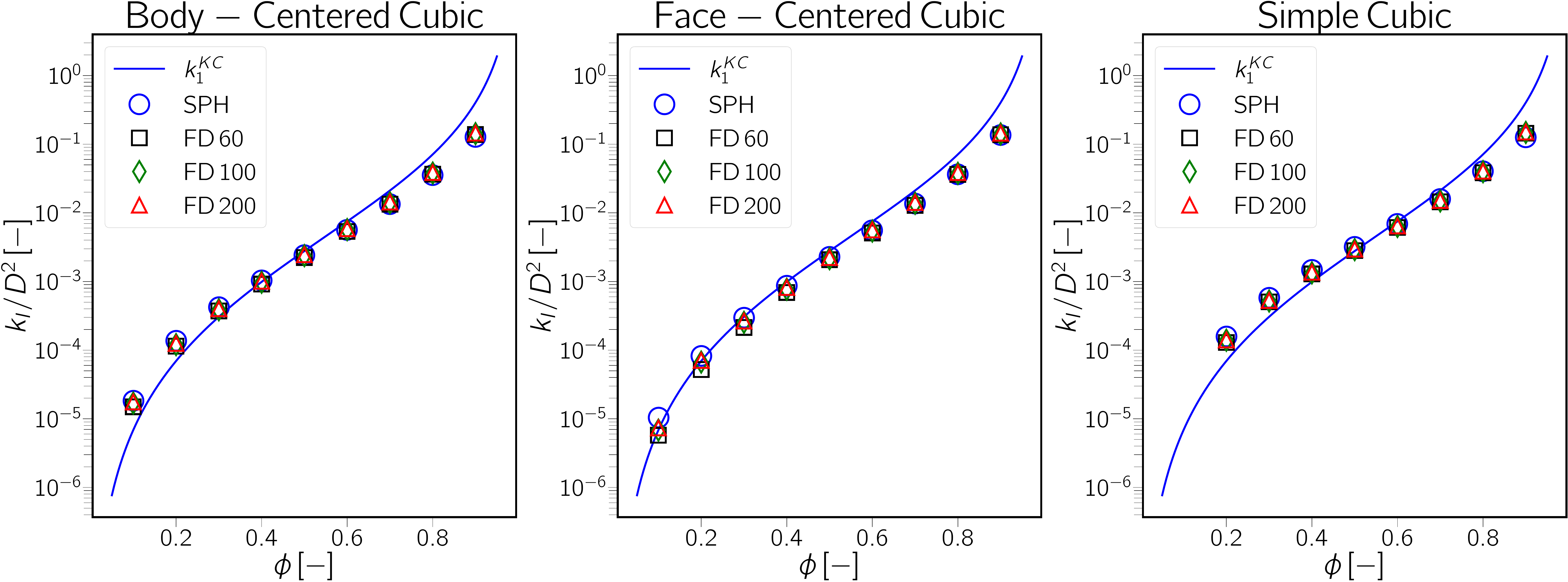}
    \caption{Comparison of Stokes solver results of normalized permeabilities with Karman-Cozeny equation and SPH solver. }
    \label{fig:spherepackings}
\end{figure}

\noindent The results of the two solvers and also in comparison with the semi-analytical solution match almost perfectly. It can be concluded that the solver is capable for a wide range of porosities. A single simulation for a resolution of $L/\mathrm{d}x = 100$ takes in average $40$ minutes on a desktop PC using $4$ cores (11th Gen Intel(R) Core(TM) i7-11700KF @ 3.60GHz).

\subsection{Permeability tensor and principal permabilities }
To compute the anisotropic permeability behavior of various materials, we must not only consider standard benchmarks and those for simple porous materials but also include benchmarks with known or adjustable principal directions. For this purpose, we create a cube with $100^3$\,voxels corresponding to $\SI{1}{\cubic\milli\meter}$ and place an ellipsoid in the center. It has the following semi-axes in Cartesian coordinates $\mathbf{e}_i$ each given in absolute voxel numbers: $\mathbf{e}_1\!: \, a = 35 ; \, \mathbf{e}_2\!: \, b = 10 ; \,\, \mathbf{e}_3\!: \, c = 35 \, .$
By using this per se periodic structure, one does not have to consider the difficulties that would arise from symmetrical or translational periodization \cite{guibert2016comparison}. We rotate the single ellipsoid around $\mathbf{e}_1$-axis in $5^\circ$ steps for $0^\circ \leq \alpha \leq 90^\circ$ and compute fluid flow and permeability, whereby the consideration of the permeability in $\mathbf{e}_1$-direction does not play a role in our evaluation for the time being. Thus the computation of the permeability tensor in Eq.(\ref{eq:3dK}) is reduced to a two-dimensional problem and eigenvalues as well as principal directions can be represented by an ellipse. 

\begin{figure}[h!]
    \centering
    \includegraphics[width=0.9\textwidth]{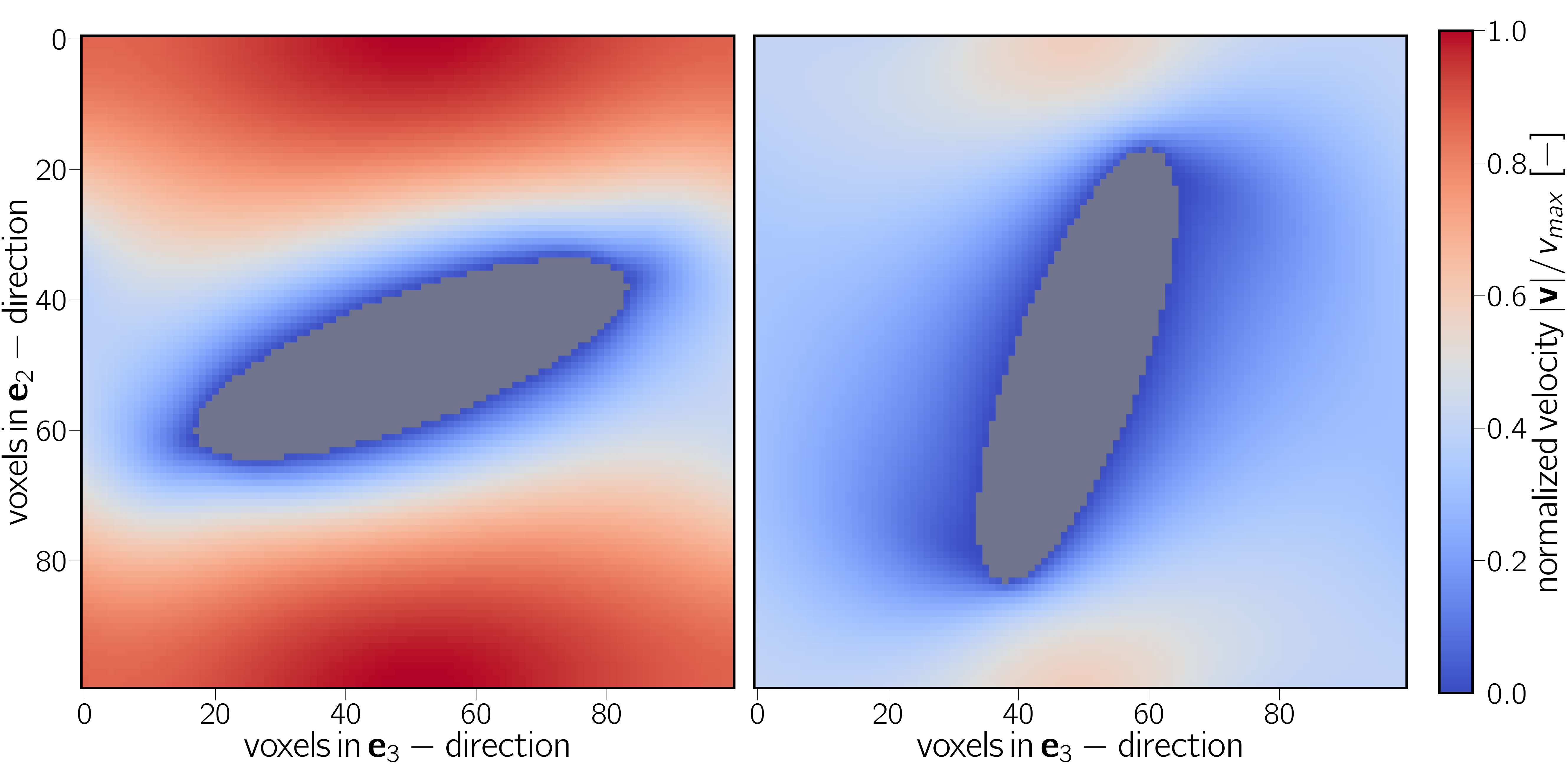}
    \caption{Normalized magnitude of velocity around an ellipsoid with $20^\circ$ (left) and $70^\circ$ inclination (right). The same pressure gradient in $\mathbf{e}_3$-direction is applied, which explains differences in the maximum velocities. Both figures show a cross-section ($\mathbf{e}_2 - \mathbf{e}_3$ plane) in the center of the domain.}
    \label{fig:velocity_ellipse}
\end{figure}

\begin{figure}[htb]
    \centering
    \includegraphics[width=0.99\textwidth]{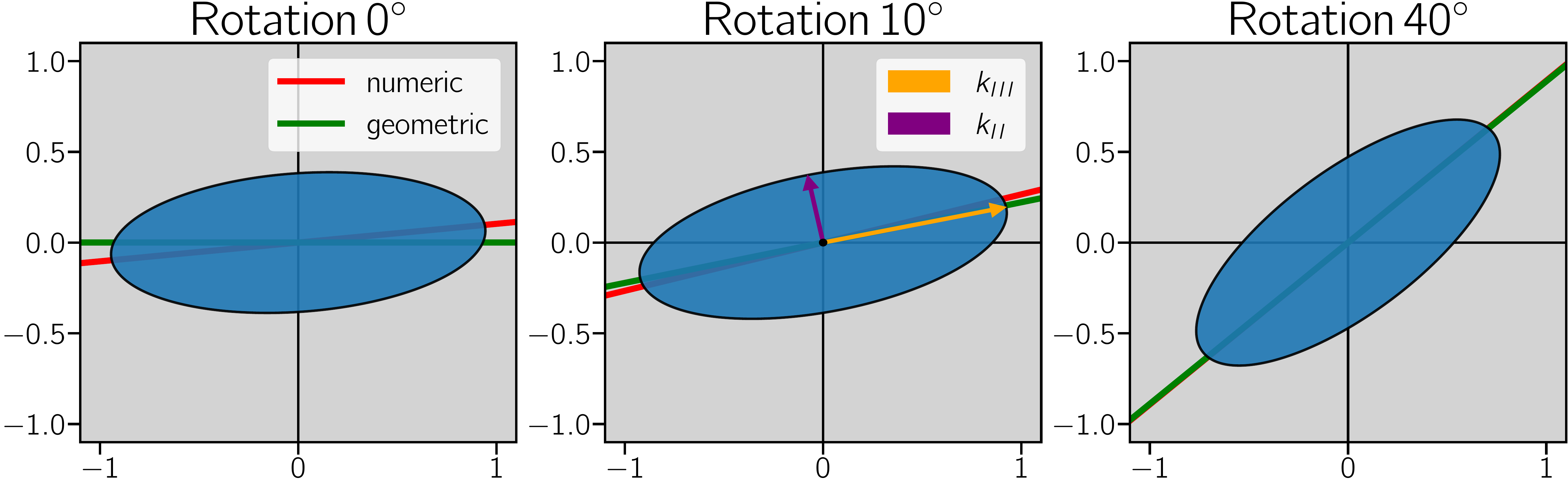}
    \caption{Plots of the coefficient matrices of the permeability tensors for rotations $ \alpha = \{0\si{\degree}, 10\si{\degree}, 40\si{\degree}\}$. Absolute values of two semi-axes of the ellipse represent the eigenvalues of the tensor. The angle of inclination (geometric information) and the computed angle of the eigenvector highlighted with a green and red line, respectively.}
    \label{fig:all_ellipsoids}
\end{figure}

\noindent Each simulation of a domain described above allows us to determine one column of the coefficient matrix $k_{ij}$, or in our simplified case two entries. The other two entries are obtained from the simulation for which the direction of the imposed pressure gradient relative to the ellipsoid is rotated by $90^\circ$. Alternative, instead of rotating the imposed pressure gradient the ellipsoid is rotated. For example, considering a rotation of $\alpha^1 = 20^\circ$, $\alpha^2 = -70^\circ$ follows. Due to symmetry, also $\alpha^2 = 70^\circ$ can be set and we get the velocity patterns, illustrated in Fig.\ref{fig:velocity_ellipse}, and compute the following coefficient matrix 

\begin{equation}\label{eq:perm_eval_elippsis}
    k_{ij} =  \begin{bmatrix}
        4.54 & 1.76 \\
        1.96 & 8.71        
\end{bmatrix}  \cdot10^{-9} \si{\meter\squared}  \approx \begin{bmatrix}
        4.54 & 1.86 \\
        1.86 & 8.71 
\end{bmatrix}  \cdot10^{-9} \si{\meter\squared}
\end{equation}
and finally receive the coefficient matrix for the principle axes system
\begin{equation}\label{eq:perm_eval_elippsis_eigen}
    \tilde{k}_{ij} =  \begin{bmatrix}
        3.82 & 0 \\
        0 & 9.42 
\end{bmatrix}  \cdot10^{-9} \si{\meter\squared}  \, .
\end{equation}
\noindent Numerically determined, one obtains slightly different values for two corresponding secondary diagonal elements. These are therefore averaged $ k_{12} = k_{21} = \frac{1}{2} (k_{12} + k_{21}) $. The same procedure would adopted for $k_{13}$, $k_{31}$ and $k_{23}$, $k_{32}$. In this reduced case,  the angle $\varphi$ between the axes in the original basis system and the principal directions is computed by
\begin{equation}
    \varphi  =  \frac{1}{2} \, \mathrm{tan}^{-1}\left(\frac{2 k_{23}}{k_{22} + k_{33}}  \right) =
    \frac{1}{2} \, \mathrm{tan}^{-1}\left(\frac{2 \cdot 1.86 \cdot10^{-9} \si{\meter\squared}}{8.71\cdot10^{-9} \si{\meter\squared} + 4.54\cdot10^{-9} \si{\meter\squared}}  \right) = 0.364 \approx 20.8\si{\degree}
\end{equation}
which matches very well to what was being specified. The listing of all values can be found in Table~\ref{tab:ellipse_allvalues} and a selection is visualized in Fig.~\ref{fig:all_ellipsoids}.

\begin{table}[h]
\footnotesize
\centering
\caption{Permeabilities, principal permeabilities and principal directions for all simulations.}
\begin{tabular}{l l l l l l l l l l l l } 
 \toprule
 $\alpha$ & $ [\, \si{\degree} \, ]$ & $0$ & $5$ & $10$ & $15$ & $20$ & $25$ & $30$ & $35$ & $40$ & $45$  \\ \midrule
 $k_{33} $ & $ [ \, \cdot 10^{-9} \si{\meter\squared} \, ]$ & $9.43$  &  $9.33$ & $9.19$ &  $8.99$ &  $8.71$ &  $8.36$ &  $7.97$ &  $7.51$ &  $7.12$ &  $6.63$ \\ 
 $k_{22} $ & $ [ \, \cdot 10^{-9} \si{\meter\squared} \, ]$ & $3.85$ &$3.92$ &$4.04$ &$4.25$ &$4.54$ &$4.91$ &$5.29$ &$5.70$ &$6.17$ &$6.63$ \\ 
 $k_{23} $ & $ [ \, \cdot 10^{-9} \si{\meter\squared} \, ]$ & $4.57$ & $0.83$ & $1.14$ & $1.50$ & $1.86$ & $2.21$ & $2.47$ & $2.59$ & $2.80$ & $2.83$ \\ 
 $k_{III} $ & $ [ \, \cdot 10^{-9} \si{\meter\squared} \, ]$  & $9.46$ & $9.45$ & $9.43$ & $9.43$ & $9.42$ & $9.44$ & $9.44$ & $9.36$ & $9.49$ & $9.46$ \\ 
 $k_{II} $ & $ [ \, \cdot 10^{-9} \si{\meter\squared} \, ]$ & $3.81$ & $3.79$ & $3.80$ & $3.81$ & $3.82$ & $3.83$ & $3.82$ & $3.86$ & $3.80$ & $3.80$ \\ 
 $\varphi $ & $ [\, \si{\degree} \, ]$ & $ 4.6 $ & $ 8.5 $ & $11.9 $ & $16.1 $ & $20.8 $ & $26.0 $ & $30.8 $ & $35.4 $ & $40.1 $ & $45.0 $ \\ \bottomrule
\end{tabular}
\label{tab:ellipse_allvalues}
\end{table}

\noindent We observe a clearly recognizable trend for increase of the permeability $k_{22}$ and decrease of $k_{33}$ with increasing rotation $\alpha$. Performing the transformation, we obtain the same eigenvalues for the permeability, except for minimal differences, which can be explained by the numerical aspects and issues related to discretization. The calculated angles for small $\alpha$ still exhibit noticeable deviations from the geometrically prescribed angles, with a maximum deviation of up to $4.6^\circ$. An error arises since the analysis for $\alpha^1 = 0\si{\degree}$ includes a simulation $\alpha^2 = 90\si{\degree}$ where the flow approaches the ellipsoid from its wide side. In Table~\ref{tab:ellipse_allvalues} we can also see two outliers for the secondary diagonal elements for $\alpha = 0\si{\degree}$ and $\alpha = 5\si{\degree}$. The larger the inclination, the more precisely the simulation result reproduces the angle presupposed in the real geometry. For $\alpha = 40\si{\degree}$ the error corresponds to only $0.1\si{\degree}$. For this comparably straightforward benchmark geometry, we can demonstrate the overall effectiveness of the workflow. It is worth highlighting that when the flow approaches the ellipsoid at $\alpha = 90\si{\degree}$, it leads to deviations in the secondary diagonal elements. This intriguingly results in enhanced solution accuracy when the system is maximally distant from one based on the principal directions.

\subsection{Testing against other codes}\label{ssec:other_codes_benchmarks}
\paragraph{Regular thin porous media (2D microfluidic devices):} 
In Sec.~\ref{ssec:dev_anios_precipitation} we investigate structures that have a small thickness compared to their lateral dimensions. The dimensions of such thin ("2D") porous materials, which are often used in microfluidic characterization of porous media flow, pose problems for numerical methods in terms of efficiency and convergence. To ensure the applicability of the solver for such requirements, we compare it with benchmarks for regular 2D porous materials described in \cite{wagner2021permeability}. The dimension of the simulated unit cell is $\SI{1}{\milli\meter} \times \SI{0.091}{\milli\meter} \times \SI{1}{\milli\meter}$ and is discretized by $250 \times 27 \times 250$\,voxels, with a solid frame of two voxels thick each, included in the $\mathbf{e}_2$ direction. All results align with the solutions provided in the benchmark paper.

\begin{table}[h!]
\footnotesize
\centering
\caption{Comparison with different types of pore scale solvers (SPH, FEM, LBM from \protect\citeA{wagner2021permeability}) for thin porous media samples and comparison of computed permeabilities.}
\begin{tabular}{ l l c c } 
 \toprule
 Radius & $\phi$ & $k_{11}$ range from \citeA{wagner2021permeability}  & $k_{11}$ (POREMAPS) \\ 
 \midrule
$\SI{0.35}{\milli\meter}$ & $0.62$  & $22.9 - 26.7 \cdot10^{-10} \, \si{\meter\squared}$ & $ 26.1 \cdot10^{-10} \, \si{\meter\squared}  $  \\ 
$\SI{0.40}{\milli\meter}$ & $0.50$  & $16.4 - 17.7 \cdot10^{-10} \, \si{\meter\squared}$ & $ 17.5 \cdot10^{-10} \, \si{\meter\squared} $   \\ 
$\SI{0.45}{\milli\meter}$ & $0.36$  & $7.54 - 8.59 \cdot10^{-10} \, \si{\meter\squared}$ & $ 8.08 \cdot10^{-10} \, \si{\meter\squared}  $\\ 
$\SI{0.47}{\milli\meter}$ & $0.31$  & $3.62 - 5.35 \cdot10^{-10} \, \si{\meter\squared}$ & $ 3.83 \cdot10^{-10} \, \si{\meter\squared} $   \\ 
$\SI{0.49}{\milli\meter}$ & $0.25$  & $0.46 - 0.54 \cdot10^{-10} \, \si{\meter\squared}$ & $ 0.39 \cdot10^{-10} \, \si{\meter\squared}$    \\ 
\bottomrule
\end{tabular}
\label{tab:sfb_domain_comparison}
\end{table}

\noindent To illustrate the performance of the solver, the SPH solver used in \citeA{wagner2021permeability} needs approximately $17$ hours on a $4$-core CPU, the presented Stokes solver, at the same resolution and on same hardware takes about $10$ minutes. 

\paragraph{Irregular sphere packings and porous rock:} 
\begin{table}[h!]
\footnotesize
\centering
\caption{Properties for different types of 3D benchmarks (range of LBM results from \protect\citeA{saxena2017references}) and comparison of computed permeabilities.}
\begin{tabular}{ l l l l l l } 
 \toprule
 Sample & $\phi$ & Voxel size & Domain size & $k_{11}$ \cite{saxena2017references}  &  $k_{11}$ (POREMAPS)  \\ [0.5ex] 
 \midrule
Sphere packing & $0.34$ & $\SI{7.0}{\micro\meter}$  & $788\times791\times793$ &  $2.438 - 2.903 \cdot10^{-10} \, \si{\meter\squared}$ & $2.512 \cdot10^{-10} \, \si{\meter\squared}$     \\ 
Berea  & $0.18$  & $\SI{2.114}{\micro\meter}$ & $1024\times1024\times1024$ &  $4.569 - 6.889\cdot10^{-13} \,\si{\meter\squared}$  & $5.772 \cdot10^{-13} \, \si{\meter\squared}$    \\ 
Fontainebleau & $0.09$ & $\SI{2.072}{\micro\meter}$ & $1024\times1024\times1024$ &  $0.642 - 1.411\cdot10^{-13} \, \si{\meter\squared}$ & $ 9.200 \cdot10^{-14} \, \si{\meter\squared}$    \\   
\bottomrule
\end{tabular}
\label{tab:saxena_bench_properties}
\end{table} 

There are several benchmark papers providing suitable geometries \cite{andra2013digital, andra2013digitalb, saxena2017references}. 
The solver is compared with different 3D benchmarks where permeabilities are computed with, among others, different LBM solvers \cite{saxena2017references}. We compute the sphere packing, one Berea sandstone (\textit{Rock1}), and one Fontainebleau sandstone (\textit{Rock3}) sample. The characteristics of the geometries and a comparison of the computed permeabilities $k_{11}$ are summarized in Table \ref{tab:saxena_bench_properties}. The results for $k_{11}$ are in accordance with the solutions determined by LBM and thus we consider the benchmarking to be succesful and completed.

\section{Applications}\label{sec:applications}
In this section, three different application scenarios are shown to demonstrate the capabilities of POREMAPS. The corresponding domain sizes and computational times are summarized in Table~\ref{tab:summary_applications}.

\begin{table}[h!]
    \centering
    \footnotesize
    \caption{Overview of computation times of considered application examples.}
    \begin{tabular}{lll}
        \toprule
         Domain size & Hardware and resources & Computation time  \\
        \midrule
        \multicolumn{3}{c}{Example 1 ``Berea'' and ``sphere packing'' (Sec.~\ref{ssec:other_codes_benchmarks} and Sec.~\ref{ssec:app_spherepacking})}\\
        \midrule
         $2048\times2048\times2048$ & SimTech Cluster, 2 Nodes & $36$ hrs. \\
         $1576\times1582\times1586$ & SimTech Cluster, 2 Nodes & $24$ hrs. \\
         $3152\times3164\times3172$ & SimTech Cluster, 4 Nodes & $\approx 80$ hrs. \\
        \midrule
        \multicolumn{3}{c}{Example 2 ``Open-cell foam ($\varepsilon_{33} = 0.0 $ and $\varepsilon_{33} = -0.5 $)'' (Sec.~\ref{ssec:app_foam}) }\\
        \midrule
         $800\times800\times1186$ & SimTech Cluster, 4 Nodes & $8$ hrs.  \\
         $800\times800\times526$ & SimTech Cluster, 4 Nodes & $3$ hrs.  \\
        \midrule
        \multicolumn{3}{c}{Example 3 ``Calcite precipitation'' (Sec.~\ref{ssec:dev_anios_precipitation})}\\
        \midrule
        $12\times1200\times1200$ & SimTech Cluster, 1 Node & 0.5--2 hrs. \\
        \bottomrule
    \end{tabular}
    \label{tab:summary_applications}
\end{table}

\subsection{Characterization of large 3D images}\label{ssec:app_spherepacking}

\begin{figure}[htb]
    \centering
    \includegraphics[width=0.5\textwidth]{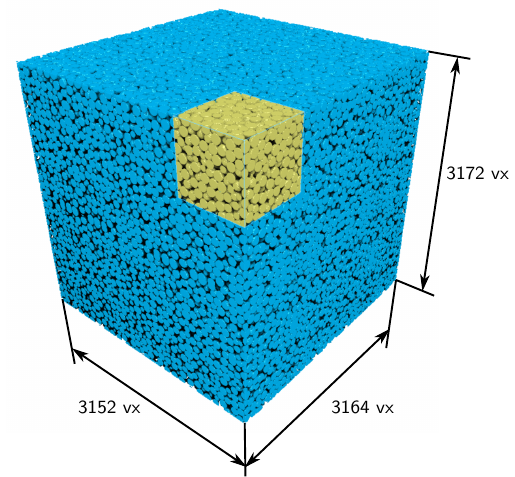}
    \caption{Illustration of the large sized computable domains using the example of a double-mirrored 3D image of a sphere packing (blue) with originally $788 \times 791 \times 793$\,voxel (yellow). The original 3D image (yellow) is taken from \protect\citeA{saxena2017references}.}
    \label{fig:example_large_domain}
\end{figure}

\noindent The solver is primarily designed to analyze large, high-resolved 3D images of porous materials typically acquired by 3D imaging methods like {\textmu}XRCT \cite{Stock2019, ruf2020open, Withers2021}. This allows the digital characterization of porous materials with regard to hydraulic permeability and can be integrated into the imaging workflow as a subsequent standard procedure. In {\textmu}XRCT imaging, resolution of REVs is often given by 3D images having more than $1000^3$\,voxel. High-resolution X-ray flat panel detectors having around $3000\times3000$\,pixel and more are not uncommon anymore. This means in terms of the reconstructed 3D images, that they can have $\approx 3000^3$\,voxel. With appropriate hardware (large memory), the presented code is capable to handle such big 3D images. 

\noindent To demonstrate this, we use the 3D image of the Berea sandstone ($1024 \times 1024 \times 1024$\,voxel), and the sphere packing ($788 \times 791 \times 793$\,voxel) from \citeA{saxena2017references}, already considered for the benchmark tests, cf. Table~\ref{tab:saxena_bench_properties}. Both data sets are mirrored in all three spatial directions resulting in $2048 \times 2048 \times 2048$\,voxel and $1576 \times 1582 \times 1586$\,voxel. The corresponding computation times are given in Table~\ref{tab:summary_applications} for a convergence criteria of $\epsilon = \frac{ \Delta |\mathbf{q}|}{|\mathbf{q}|} < 10^{-6}$. Latter data set is mirrored a second time in each direction resulting in $3152 \times 3164 \times 3172$ voxel (equals $64$ times the original geometry), cf. Figure~\ref{fig:example_large_domain}. Due to parallelized IO routines, this can be implemented using $4$ nodes, with a total of $512$ cores, requiring a total of $\approx 1.5\,\mathrm{TB}$ RAM. 

\subsection{Deformation-dependent permeability and permeability anisotropy -- uniaxial compression of an open-cell foam}\label{ssec:app_foam}

The deformation-dependent permeability of open-cell foams has been studied experimentally by several researches. Different theoretical models were proposed to predict the effect of strain on permeability, cf. \citeA{Dawson2007, Markert2007} and therein cited literature. In experimental studies, often uniaxial compression loading is imposed on a foam sample and the deformation-depending permeability is measured in one direction, often in the same direction as the imposed load. Measuring the permeability in different directions is technically challenging. Using non-destructive {\textmu}XRCT imaging is a possible approach to overcome  this problem when the fluid-solid interaction is negligible and the foam can be considered as rigid at a given deformation state. In this case, the foam structure is imaged under different loading conditions in 3D. Based on the time-series of 3D images followed by a subsequent post-processing provides the possibility to perform virtual experiments and to characterize the structure and the deformation-dependency in more detail. 

\begin{figure}[htb]
    \centering
    \includegraphics[width=1.0\textwidth]{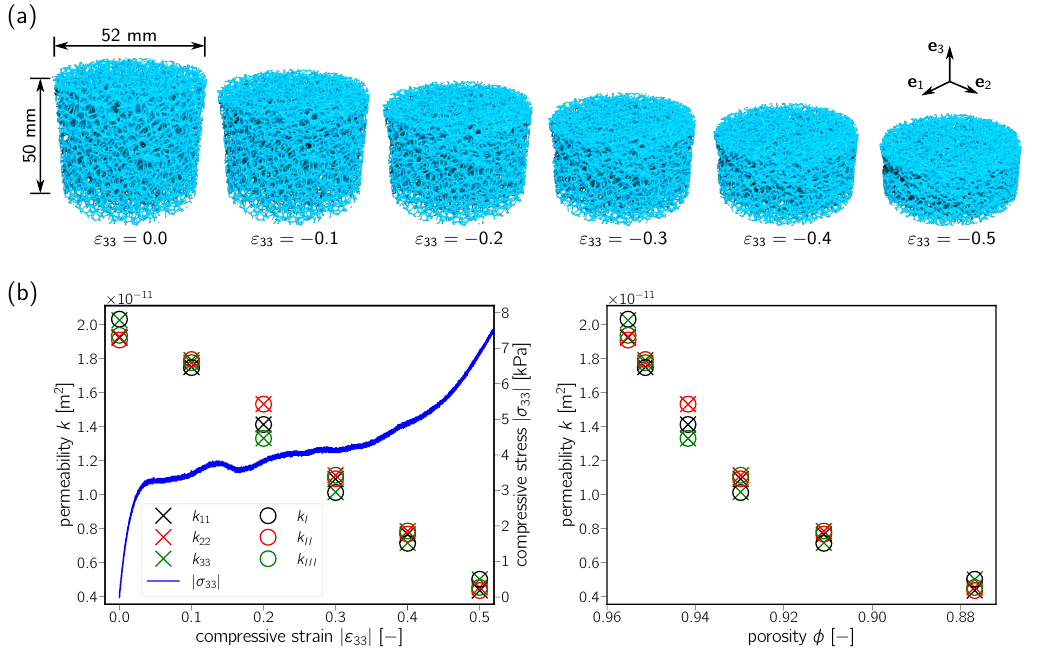}
    \caption{Deformation-dependent permeability and anisotropy of an open-cell PUR foam sample (10\,PPI) based on {\textmu}XRCT images \protect\cite{ruf2023visualization, darus-3010_2024}. (a) Segmented {\textmu}XRCT images for different uniaxial compression loading states. (b) Determined permeabilities in $\mathbf{e}_1$-, $\mathbf{e}_2$-, $\mathbf{e}_3$-direction and principle permeabilities.}
    \label{fig:permeabilities_foams}
\end{figure}

\noindent This is exemplary shown for an open-cell Polyurethane (PUR) cylindrical foam sample with \mbox{10\,PPI} subjected to uniaxial compressive loading and imaged at discrete loading states (engineering compressive strain $\varepsilon_{33} = -\{0.0, 0.1, 0.2, 0.3, 0.4, 0.5 \}$), see Figure~\ref{fig:permeabilities_foams}(a). For the imaging, the system presented in \citeA{ruf2020open,ruf2023RSI} was employed. For the series of 3D images, the permeability tensor is determined using cubic subvolumes of size $800\times800\times$ $526$--$1186$\,voxel with a uniform voxel size of \mbox{74.8\,{\textmu}m}. The permeability tensor (three simulations) for each loading condition is determined in \mbox{$\approx 3$-$8$\,hrs.}. The permeability in axial direction ($\mathbf{e}_3$) and radial directions ($\mathbf{e}_1$, $\mathbf{e}_2$) and the principle values are shown in Figure~\ref{fig:permeabilities_foams}(b) over the applied strain (left) and the foam porosity (right). At all deformation states, the radial permeabilities $k_{11}$ and $k_{22}$ are quite similar and differ slightly from the axial permeability $k_{33}$ of the cylindrical sample. In general, it can be said that the permeability anisotropy does not correlate in a systematic way, neither with the strain nor with the porosity, which contradicts the expectation.

\subsection{Anisotropy development during calcite precipitiation}\label{ssec:dev_anios_precipitation}

\noindent Predicting pore-scale clogging phenomena in heterogeneous porous materials presents a significant challenge. These processes can occur inadvertently, and if they cannot be prevented, there is a need to manage them, such as in the case of clogged filters. Conversely, these processes might also be intentional, such as in the case of blocking subsurface cracks. We focus our investigation on a specific scenario where pore spaces gradually become obstructed due to a chemically induced precipitation process. The experimental data set, \citeA{darus-1799_2022}, serves as the basis for our study, and the details of data acquisition and the experimental arrangement are elaborated upon in \citeA{weinhardt2021experimental, weinhardt2022porosity, weinhardt2022spatiotemporal}. We have already described this procedure of investigation on a smaller extent as a proof of concept in \citeA{krach2023comparing}. 

\begin{figure}[h!]
    \centering
    \includegraphics[width=0.9\textwidth]{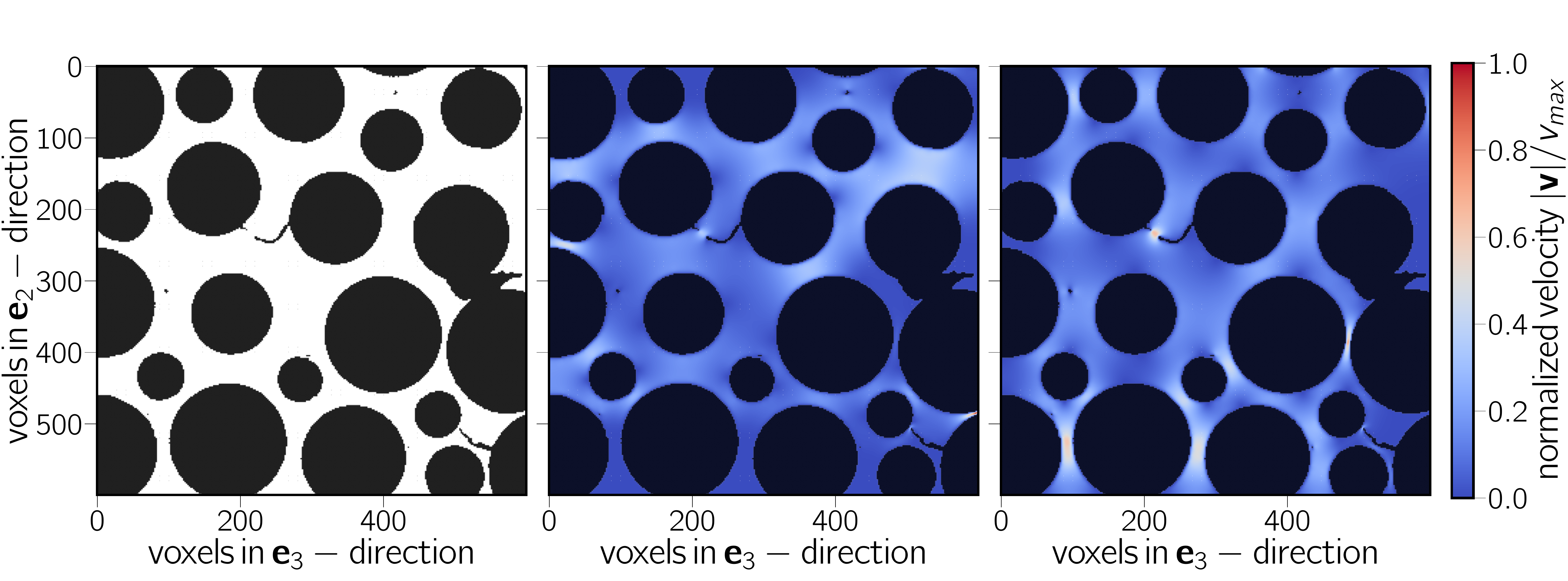}
    \includegraphics[width=0.9\textwidth]{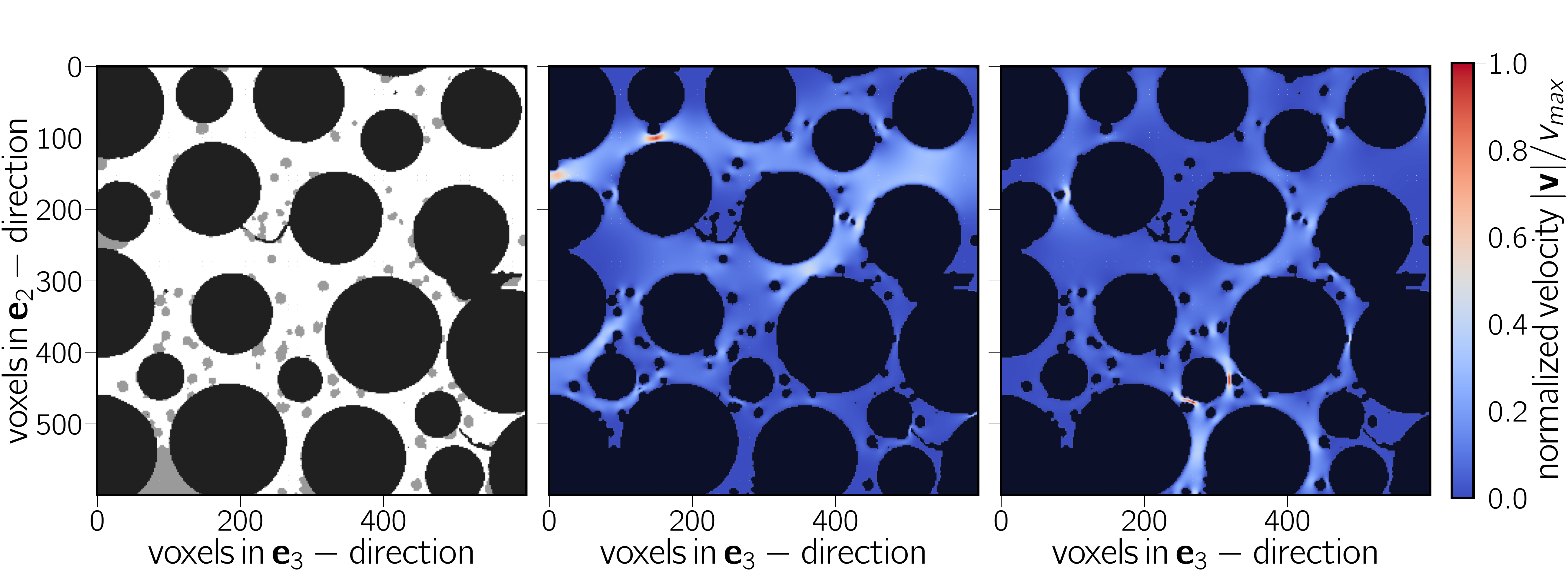}
    \includegraphics[width=0.9\textwidth]{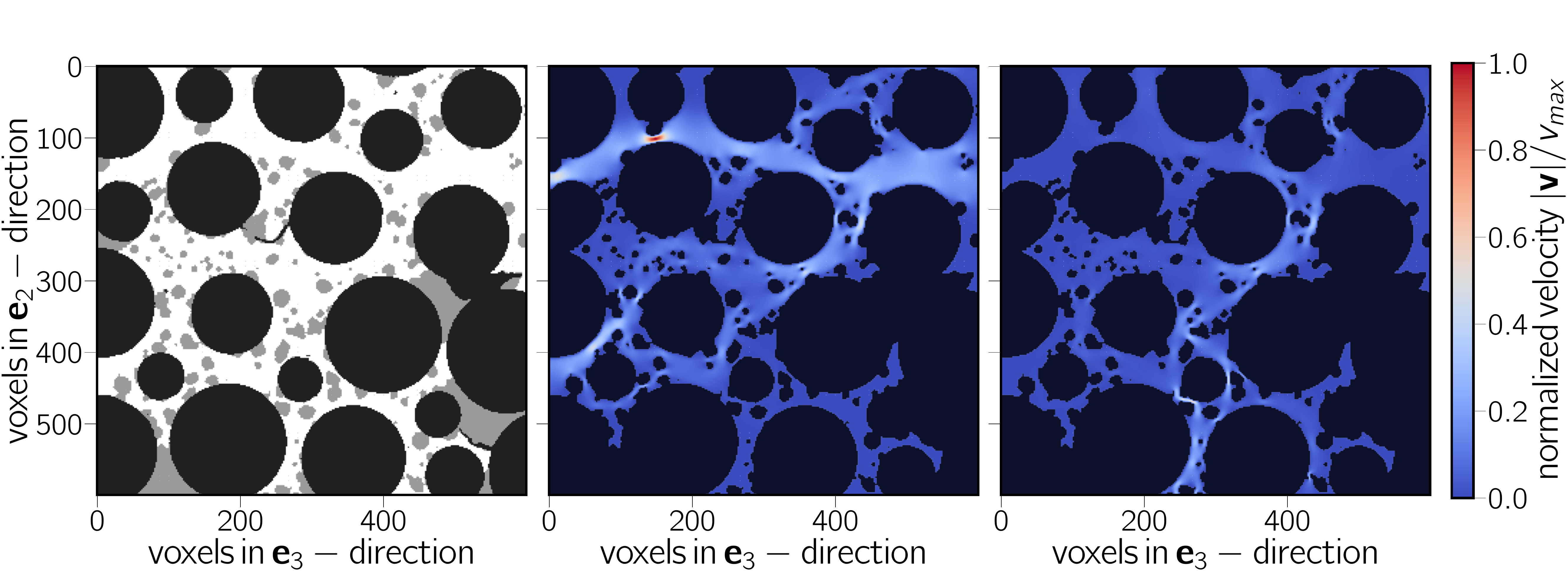}
    \includegraphics[width=0.9\textwidth]{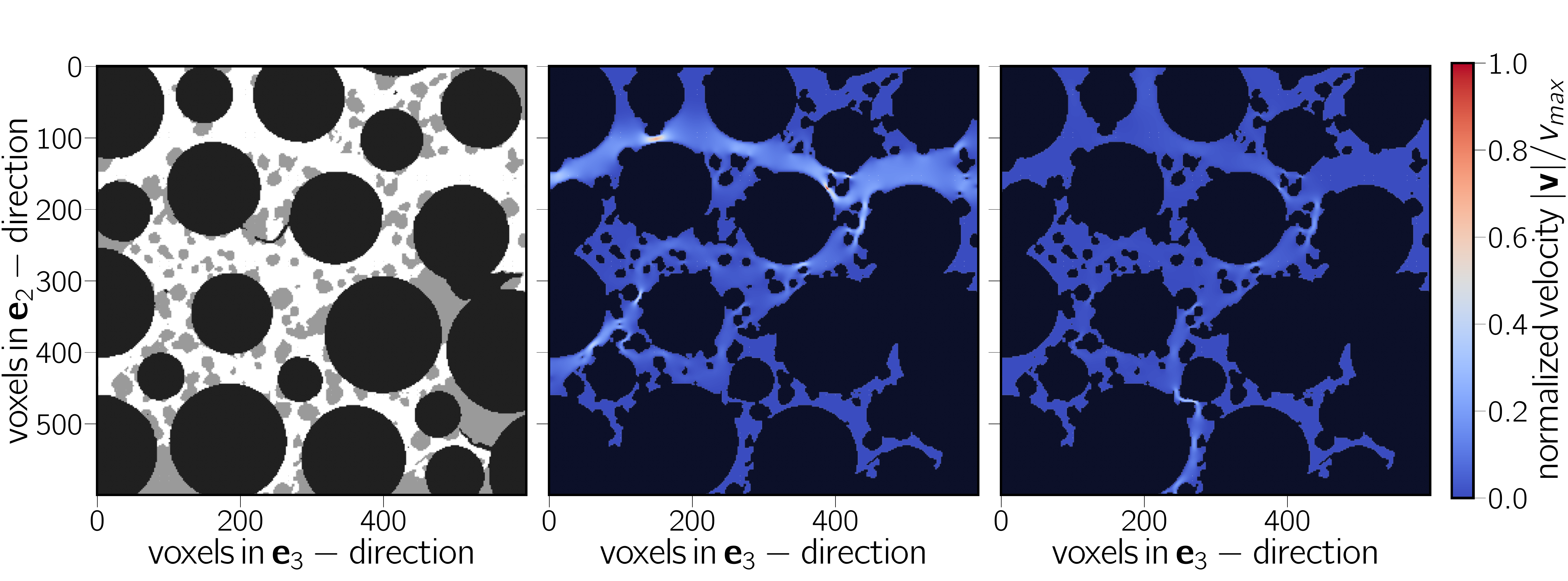}
    \caption{Geometry and flow patterns at different times (increasing from top to bottom) during the experiment. The left column represents the input domain (black - solid columns, gray - precipitate). For the simulations, the precipitates are attributed to the solid. Column $2$ and $3$ show the normalized absolute velocities based on pressure gradients in $\mathbf{e}_3$- and $\mathbf{e}_2$-direction.}
    \label{fig:flow_path_dev}
\end{figure}

\noindent Unlike the experimental procedure, we can determine the permeability $k_{33}$ in the $\mathbf{e}_3$-direction (aligned with the pressure gradient in the experiment) and the permeability $k_{22}$ in the $\mathbf{e}_2$-direction (perpendicular to the pressure gradient) at various time steps during the experiment. This investigation includes $137$ individual simulations, where for $57$ time steps the domain is percolating in both directions, resulting in $2$ simulations per time step. In $\mathbf{e}_3$-direction a flow path remains unblocked for a longer time and $80$ time steps are investigated. A simulation ($12 \times 1200 \times 1200$\,voxel) typically runs for an average of $2$ hours on a node using $121$ CPU-cores. We explicitly fix the domain decomposition to $1\times11\times11$ ranks.  

\begin{figure}
    \centering
    \includegraphics[width=0.9\textwidth]{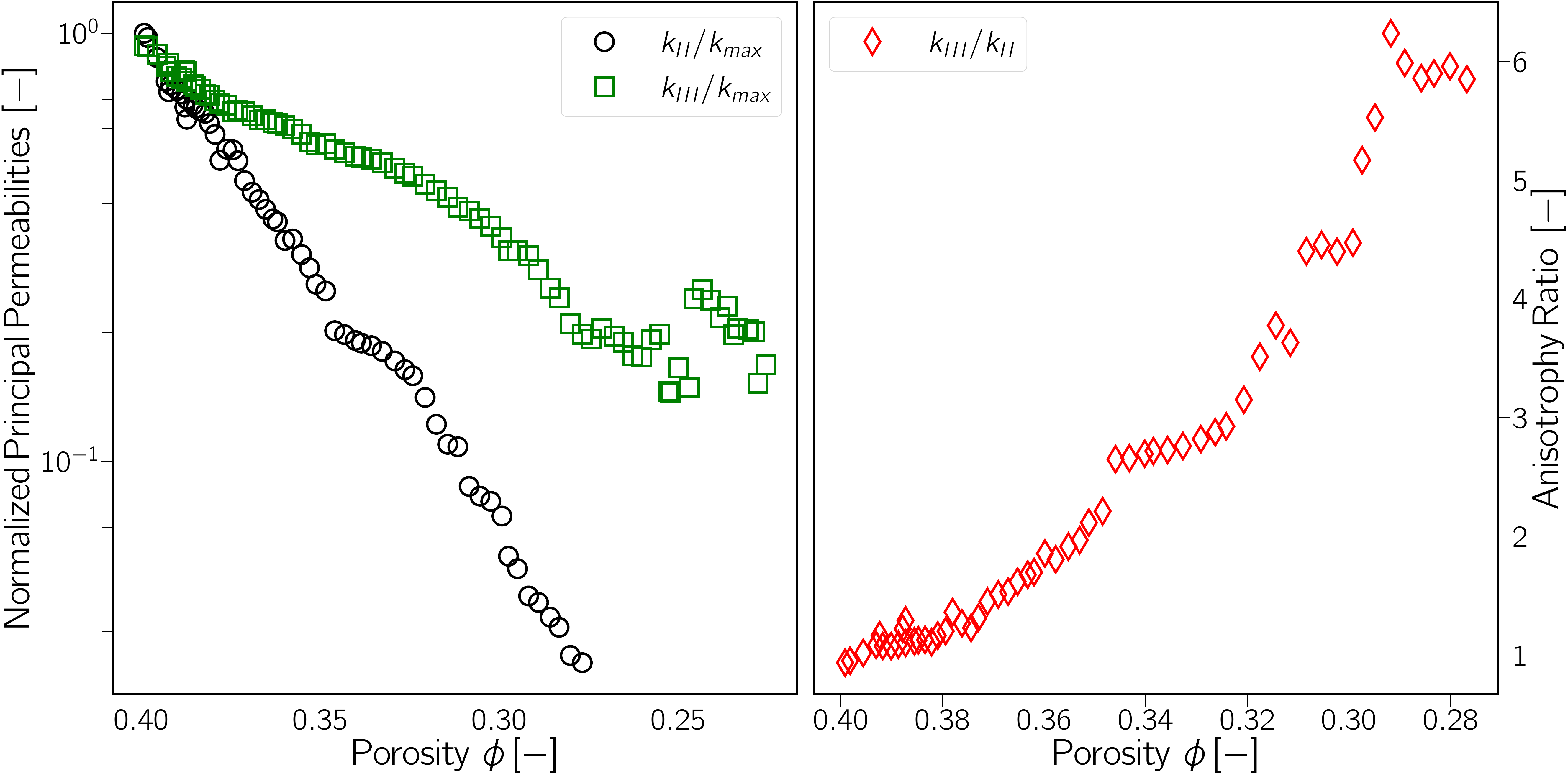}
    \caption{Normalized eigenvalues of the permeability tensor ($k_{II}$, $k_{III}$) over the course of a decreasing porosity (left) and the relation of both (${k_{III}}/{k_{II}}$) illustrating a clear development of anisotropic hydraulic properties (right). }
    \label{fig:porosity_permeability_relationship}
\end{figure}

\noindent 2D image data is used as basis and replicated in the third spatial direction to simulate the original dimensions of the microfluidic experiment in 3D. We are particularly interested in a preferential flow path that forms in the upper region of the domain and stays almost free during the experiment. Accordingly a subdomain is extracted. In the course of the experiment, more and more precipitate is accumulated in the domain, thus increasing the solid fraction and decreasing porosity and permeability. Fig.~\ref{fig:flow_path_dev} shows the geometries (black - solid body, gray - precipitate) and the velocity patterns. Note that we simulate the same domain twice with pressure gradient from left to right ($\mathbf{e}_3$-direction: Fig.~\ref{fig:flow_path_dev}, center column) and pressure gradient from top to bottom ($\mathbf{e}_2$-direction: Fig.~\ref{fig:flow_path_dev}, right column). The simulations include all time steps of the experiment up to the point where there is no flow path through the porous material at all. The material clogs first in the $\mathbf{e}_2$-direction, at a porosity of $\phi = 27.7\, \%$, while at this time a flow path remains open in the $\mathbf{e}_3$-direction. 

\noindent The focus of the analysis lies on the development of the degree of anisotropy in the course of the experiment. For this purpose, the eigenvalues of the $2 \times 2$ permeability tensor are determined. Fig.~\ref{fig:porosity_permeability_relationship} shows the eigenvalues $k_{II}$ and $k_{III}$ plotted against the porosity $\phi$. The principal permeability $k_{II}$ decreases significantly faster than $k_{III}$. The degree of anisotropy, the ratio of the eigenvalues $\frac{k_{III}}{k_{II}}$ (Fig.~\ref{fig:porosity_permeability_relationship}, right) increases stepwise linear in the course of the experiment up to a ration of $6$. Thus, this analysis provides an additional benefit that experiments cannot provide. 

\section{Conclusion}
A FDM solver, POREMAPS, for the incompressible Stokes equations using the artificially compressible method was presented to compute the permeability tensor $\mathbf{k}$ of various porous materials. Different benchmarks have been performed, which demonstrate that the solver can handle a variety of problems for a range of porosities swiftly and robustly. We present two compelling applications that perfectly align with the capabilities of the solver, demonstrating its versatility in tackling complex scenarios that demand extensive simulations. The first application focuses on open-cell foams, where we establish that any anisotropy in permeability is unlikely to develop during deformation up to $\varepsilon_{33} = -0.5$. Our findings underscore the absence of permeability directionality, what we would have expected otherwise. The second application addresses plugging porous materials, characterized by a distinct preferred orientation. In this context, we not only quantify the resulting anisotropy but also offer insights that highlight the solvers key advantages and experimental constraints.

\noindent While acknowledging the computational time and hardware prerequisites, it's worth noting that they remain modest when contrasted with alternative methodologies. The solver's user-friendly nature, coupled with its direct compatibility with experimental imaging data, opens up promising avenues in various domains.

\section*{Acknowledgement}
D.K. and H.S. acknowledge funding from the Deutsche Forschungsgemeinschaft (DFG, German Research Foundation) under Germany's Excellence Strategy - EXC 2075 – 390740016. M.R. and H.S. acknowledge funding from the DFG through the project STE 969/13-1 (Project No. 357361983) and supporting this work within the SFB 1313 (Project No. 327154368). We acknowledge the support by the Stuttgart Center for Simulation Science (SC SimTech).

\section*{Conflicts of Interest}
The authors declare that they have no known competing financial interests or personal relationships that could have appeared to influence the work reported in this paper. 

\section*{Author Contributions}
D. Krach: Conceptualization, Methodology, Investigation, Software, Validation, Data Curation, Writing - Original Draft, Visualization, Project administration. M. Ruf: Investigation, Software, Data Curation, Writing - Original Draft, Writing - Review \& Editing, Visualization. H. Steeb: Conceptualization, Writing - Review \& Editing, Supervision, Funding acquisition.

\section*{Data, Code \& Protocol Availability}
The complete source code including input data of the shown application examples that support the findings of this study is openly available in the Data Repository of the University of Stuttgart (DaRUS) at \url{https://doi.org/10.18419/darus-3676} \cite{krach2024a}.

\appendix
\section{Nondimensional balance equations:}\label{appendix:A}
\paragraph{Nondimensional quantities and operators}
We introduce the dimensionless length $l^\ast$, velocity $\mathbf{v}^\ast$, and time 
$t^\ast$ by the ratio of physical property divided by characteristic property ($\mathscr{L}$, $\mathscr{V}$, $T=\mathscr{L}/\mathscr{V}$)
\begin{equation}
    l^\ast  = \frac{l}{\mathscr{L}} \quad \rightarrow \quad l = l^\ast \mathscr{L}\, , \qquad \mathbf{v}^\ast = \frac{\mathbf{v}}{\mathscr{V}}  \quad \rightarrow \quad \mathbf{v} =  \mathbf{v}^\ast \mathscr{V} \, \quad \text{and} \quad t^\ast = \frac{t}{T} \quad \rightarrow \quad t = \frac{t^\ast \mathscr{L}}{\mathscr{V}} \, .
\end{equation}

The nondimensional differential operators $\overset{\ast}{\mathrm{grad}} (\bullet)$ and $\overset{\ast}{\mathrm{div}} (\bullet)$ and time derivative $\overset{\ast}{(\bullet)}\!\!^{\boldsymbol{\cdot}}$ are given as follows:
\begin{equation}\label{eq:dimless_op}
    \begin{aligned}
    \overset{\ast}{\mathrm{grad}} (\bullet) = \mathscr{L} \,\mathrm{grad} (\bullet) \quad & \rightarrow  \quad \mathrm{grad} (\bullet) = \frac{1}{\mathscr{L}} \overset{\ast}{\mathrm{grad}} (\bullet) \, , \qquad
    \overset{\ast}{\mathrm{div}} (\bullet) = \mathscr{L} \, \mathrm{div} (\bullet) \quad  \rightarrow  \quad \mathrm{div} (\bullet) = \frac{1}{\mathscr{L}} \overset{\ast}{\mathrm{div}} (\bullet) \\
    \overset{\ast}{(\bullet)}\!^{\boldsymbol{\cdot}} = T (\bullet)^{\boldsymbol{\cdot}} = \frac{\mathscr{L}}{\mathscr{V}} (\bullet)^{\boldsymbol{\cdot}} \quad & \rightarrow  \quad (\bullet)^{\boldsymbol{\cdot}} = \frac{\mathscr{V}}{\mathscr{L}} \overset{\ast}{(\bullet)}\!^{\boldsymbol{\cdot}} \, .
    \end{aligned}
\end{equation}
This allows for the selection of the pressure and density scale
\begin{equation}
    \rho^\ast = \frac{\rho}{\rho_0} \quad \rightarrow \quad \rho = \rho_0 \rho^\ast  \, \quad \text{and} \quad p^\ast = \frac{p}{\rho_0 \mathscr{V}^2} \quad \rightarrow \quad p = p^\ast \rho_0 \mathscr{V}^2 \, ,
\end{equation}
where $\rho_0$ is the rest density of the fluid. 

\paragraph{Nondimensional governing equations}
For the mutual dependence of pressure $p$ and density $\rho$ we employ the  constitutive equation and its derivative
\begin{equation}\label{eq:constitutive_pressure}
    \rho = \rho_0 \, \left[ \frac{p}{K^F} + 1 \right] \quad \rightarrow  \quad  \dot{\rho} = \rho_0 \, \frac{\dot{p}}{K^F} \, .
\end{equation}
Artificial compressibility for pseudo-unsteady Stokes equations is introduced by considering the linearized balance of mass (Eq.~(\ref{eq:mass_balance})) for a compressible fluid
\begin{equation}\label{eq:lin_conti}
    \dot{\rho} + \rho_0 \, \mathrm{div} \, \mathbf{v} = 0 \, ,
\end{equation}
where for steady state $\dot{\rho} \to 0$ holds. The Stokes equations (Eq.~(\ref{eq:stokes_equtaions_physical})) are analogously extended
\begin{equation}\label{eq:pseudo_unsteady_stokes}
    \rho \mathbf{\dot{v}} = \mu \, \mathrm{div}(\mathrm{grad} \, \mathbf{v}) - \mathrm{grad}\, p
\end{equation}
where $\rho \mathbf{\dot{v}} \to 0$ holds for steady state. We apply the dimensionless divergence operator, dimensionless velocity and density to Eq.~(\ref{eq:lin_conti}) and receive
\begin{equation}\label{eq:dimless_mass_balance_derivation}
    \frac{\mathscr{V}}{\mathscr{L}} \overset{\ast}{(\rho_0 \rho^\ast)}\!^{\boldsymbol{\cdot}} + \frac{\rho_0}{\mathscr{L}} \, \overset{\ast}{\mathrm{div}} (\mathbf{v}^\ast \mathscr{V}) = 0 \quad  \rightarrow  \quad \overset{\ast}{(\rho^\ast)}\!^{\boldsymbol{\cdot}} + \overset{\ast}{\mathrm{div}} (\mathbf{v}^\ast) = 0 \, .
\end{equation}
Using the speed of sound for a compression-type wave $c = \sqrt{\frac{K^F}{\rho_0}}$ with its dimensionless form $c = c^\ast \mathscr{V}$ the constitutive equation for the pressure Eq.~(\ref{eq:constitutive_pressure}) is transformed in the same way, yielding
\begin{equation}\label{eq:pressure_density_relation}
    \dot{\rho} = \rho_0 \, \frac{\dot{p}}{K^F} \quad \rightarrow \quad \dot{\rho} = \frac{\dot{p}}{c^2} \quad \rightarrow \quad \frac{\mathscr{V}}{\mathscr{L}} \overset{\ast}{(\rho_0 \rho^\ast)}\!^{\boldsymbol{\cdot}} = \frac{\mathscr{V}  \overset{\ast}{(p^\ast \rho_0 \mathscr{V}^2)}\!^{\boldsymbol{\cdot}}}{\mathscr{L} \, {c^\ast}^2 \mathscr{V}^2} \quad \rightarrow \quad \overset{\ast}{(\rho^\ast)}\!^{\boldsymbol{\cdot}} = \frac{\overset{\ast}{(p^\ast)}\!^{\boldsymbol{\cdot}}}{{c^\ast}^2}
\end{equation}
Plugging Eq.~(\ref{eq:pressure_density_relation}) in Eq.~(\ref{eq:dimless_mass_balance_derivation})  the nondimensional balance of mass is obtained
\begin{equation}
    \overset{\ast}{(p^\ast)}\!^{\boldsymbol{\cdot}} = - {c^\ast}^2 \overset{\ast}{\mathrm{div}} (\mathbf{v}^\ast) \, .
\end{equation}
The nondimensional form of the pseudo-unsteady Stokes problem (Eq.~(\ref{eq:pseudo_unsteady_stokes})) yields 
\begin{equation}
    \frac{\rho_0 \rho^\ast \mathscr{V}}{\mathscr{L}} \overset{\ast}{(\mathbf{v}^\ast \mathscr{V})}\!^{\boldsymbol{\cdot}} = \frac{\mu}{\mathscr{L}^2} \, \overset{\ast}{\mathrm{div}} \, \overset{\ast}{\mathrm{grad}} (\mathbf{v}^\ast \mathscr{V}) - \frac{1}{\mathscr{L}} \overset{\ast}{\mathrm{grad}} (p^\ast \rho_0 \mathscr{V}^2) \\ \quad \rightarrow \quad \rho^\ast \overset{\ast}{(\mathbf{v}^\ast )}\!^{\boldsymbol{\cdot}} = \frac{1}{\mathrm{Re}} \, \overset{\ast}{\mathrm{div}} \, \overset{\ast}{\mathrm{grad}} \, (\mathbf{v}^\ast) - \overset{\ast}{\mathrm{grad}} (p^\ast) \, .
\end{equation}

\section{Implementation of fluid-solid boundary conditions}\label{appendix:B}
%
The Taylor expansion around the point $x_0$ is generally given by the following series:
\begin{align}\label{eq:taylor_series}
	f(x)= \sum_{n=0}^{\infty} \frac{f^{(n)}(x_0)}{n!}(x-x_0)^n =
		  f^{(0)} + f^{(1)}(x-x_0)^1 + \frac{1}{2}f^{(2)}(x-x_0)^2 + \cdots \, .
\end{align}
As briefly described in Sec.~\ref{sec:FD} for case 2 depicted in Fig.~\ref{fig:neighborhoodCases}, for the other given cases following system of equations can be derived. By solving the given sets of equation systems, the corresponding second-order derivatives ($\frac{\partial^2 v_k}{\partial x_i^2}$) in the perpendicular directions to the considered flow velocity direction for $i,j,k =  (1,2,3), (2,3,1), (3,1,2)$ can be determined.

\noindent Case 1:
\begin{align}
	\left.\begin{aligned}
	v_{k_1} &\approx v_{k_2} - \frac{1}{2} \frac{\partial v_k}{\partial x_i} + \frac{1}{8} \frac{\partial^2 v_k}{\partial x_i^2} - \frac{1}{48} \frac{\partial^3 v_k}{\partial x_i^3} = 0 \\
	v_{k_3} &\approx v_{k_2} + \frac{1}{2} \frac{\partial v_k}{\partial x_i} + \frac{1}{8} \frac{\partial^2 v_k}{\partial x_i^2} + \frac{1}{48} \frac{\partial^3 v_k}{\partial x_i^3} = 0
	\end{aligned}
	\right\}
\quad \leadsto \frac{\partial^2 v_k}{\partial x_i^2} = -8 v_{k_2}
\end{align}

\noindent  Case 2:
\begin{align}
	\left.\begin{aligned}
	v_{k_1} &\approx v_{k_2} - \frac{1}{2} \frac{\partial v_k}{\partial x_i} + \frac{1}{8} \frac{\partial^2 v_k}{\partial x_i^2} - \frac{1}{48} \frac{\partial^3 v_k}{\partial x_i^3} = 0 \\
	v_{k_3} &\approx v_{k_2} + \frac{\partial v_k}{\partial x_i} + \frac{1}{2} \frac{\partial^2 v_k}{\partial x_i^2} + \frac{1}{6} \frac{\partial^3 v_k}{\partial x_i^3} \\
	v_{k_4} &\approx v_{k_2} + \frac{3}{2} \frac{\partial v_k}{\partial x_i} + \frac{9}{8} \frac{\partial^2 v_k}{\partial x_i^2} + \frac{9}{16} \frac{\partial^3 v_k}{\partial x_i^3} = 0
	\end{aligned}
	\right\}
\quad \leadsto \frac{\partial^2 v_k}{\partial x_i^2} = \frac{8}{3} v_{k_3} - \frac{16}{3} v_{k_2}
\end{align}

\noindent Case 3:
\begin{align}
	\left.\begin{aligned}
	v_{k_1} &\approx v_{k_3} - \frac{3}{2} \frac{\partial v_k}{\partial x_i} + \frac{9}{8} \frac{\partial^2 v_k}{\partial x_i^2} - \frac{9}{16} \frac{\partial^3 v_k}{\partial x_i^3} = 0 \\
	v_{k_2} &\approx v_{k_3} - \frac{\partial v_k}{\partial x_i} + \frac{1}{2} \frac{\partial^2 v_k}{\partial x_i^2} - \frac{1}{6} \frac{\partial^3 v_k}{\partial x_i^3} \\
	v_{k_4} &\approx v_{k_3} + \frac{1}{2} \frac{\partial v_k}{\partial x_i} + \frac{1}{8} \frac{\partial^2 v_k}{\partial x_i^2} + \frac{1}{48} \frac{\partial^3 v_k}{\partial x_i^3} = 0
	\end{aligned}
	\right\}
\quad \leadsto \frac{\partial^2 v_k}{\partial x_i^2} = \frac{8}{3} v_{k_2} - \frac{16}{3} v_{k_3}
\end{align}

\noindent Case 4:
\begin{align}
	\left.\begin{aligned}
	v_{k_1} &\approx v_{k_2} - \frac{1}{2} \frac{\partial v_k}{\partial x_i} + \frac{1}{8} \frac{\partial^2 v_k}{\partial x_i^2} - \frac{1}{48} \frac{\partial^3 v_k}{\partial x_i^3} = 0 \\
	v_{k_3} &\approx v_{k_2} + \frac{\partial v_k}{\partial x_i} + \frac{1}{2} \frac{\partial^2 v_k}{\partial x_i^2} + \frac{1}{6} \frac{\partial^3 v_k}{\partial x_i^3} \\
	v_{k_4} &\approx v_{k_2} + 2 \frac{\partial v_k}{\partial x_i} + 2 \frac{\partial^2 v_k}{\partial x_i^2} + \frac{8}{6} \frac{\partial^3 v_k}{\partial x_i^3}
	\end{aligned}
	\right\}
\quad \leadsto \frac{\partial^2 v_k}{\partial x_i^2} = 2 v_{k_3} - 5 v_{k_2} - \frac{1}{5} v_{k_4}
\end{align}

\noindent Case 5:
\begin{align}
	\left.\begin{aligned}
	v_{k_1} &\approx v_{k_3} - 2 \frac{\partial v_k}{\partial x_i} + 2 \frac{\partial^2 v_k}{\partial x_i^2} - \frac{8}{6} \frac{\partial^3 v_k}{\partial x_i^3} \\
	v_{k_2} &\approx v_{k_3} - \frac{\partial v_k}{\partial x_i} + \frac{1}{2} \frac{\partial^2 v_k}{\partial x_i^2} - \frac{1}{6} \frac{\partial^3 v_k}{\partial x_i^3} \\
	v_{k_4} &\approx v_{k_3} + \frac{1}{2} \frac{\partial v_k}{\partial x_i} + \frac{1}{8} \frac{\partial^2 v_k}{\partial x_i^2} + \frac{1}{48} \frac{\partial^3 v_k}{\partial x_i^3} = 0
	\end{aligned}
	\right\}
\quad \leadsto \frac{\partial^2 v_k}{\partial x_i^2} = 2 v_{k_2} - \frac{1}{5} v_{k_1} - 5 v_{k_3}
\end{align}

\noindent Case 6:
\begin{align}
	\left.\begin{aligned}
	v_{k_1} &\approx v_{k_2} - \frac{\partial v_k}{\partial x_i} + \frac{1}{2} \frac{\partial^2 v_k}{\partial x_i^2} - \frac{1}{6} \frac{\partial^3 v_k}{\partial x_i^3} \\
	v_{k_3} &\approx v_{k_2} + \frac{\partial v_k}{\partial x_i} + \frac{1}{2} \frac{\partial^2 v_k}{\partial x_i^2} + \frac{1}{6} \frac{\partial^3 v_k}{\partial x_i^3}
	\end{aligned}
	\right\}
\quad \leadsto \frac{\partial^2 v_k}{\partial x_i^2} = v_{k_1} - 2 v_{k_2} + v_{k_3}
\end{align}

\noindent Case 7 (4b):
\begin{align}
	\left.\begin{aligned}
	v_{k_1} &\approx v_{k_2} - \frac{1}{2}\frac{\partial v_k}{\partial x_i} + \frac{1}{8} \frac{\partial^2 v_k}{\partial x_i^2} = 0 \\
	v_{k_3} &\approx v_{k_2} + \frac{\partial v_k}{\partial x_i} + \frac{1}{2} \frac{\partial^2 v_k}{\partial x_i^2}
	\end{aligned}
	\right\}
\quad \leadsto \frac{\partial^2 v_k}{\partial x_i^2} = \frac{4}{3} v_{k_3} - 4 v_{k_2}
\end{align}

\noindent Case 8 (5b):
\begin{align}
	\left.\begin{aligned}
	v_{k_1} &\approx v_{k_2} - \frac{\partial v_k}{\partial x_i} + \frac{1}{2} \frac{\partial^2 v_k}{\partial x_i^2} \\
	v_{k_3} &\approx v_{k_2} + \frac{1}{2}\frac{\partial v_k}{\partial x_i} + \frac{1}{8} \frac{\partial^2 v_k}{\partial x_i^2} = 0
	\end{aligned}
	\right\}
\quad \leadsto \frac{\partial^2 v_k}{\partial x_i^2} = \frac{4}{3} v_{k_1} - 4 v_{k_2}
\end{align}

\section{Permeability computation}\label{appendix:C}
For $\mathrm{Re} < 1.0$ the entries of the permeability tensor can be computed using Darcy`s law \cite{darcy1856fontaines} 
\begin{equation}\label{eq:darcy_appendix}
    \mathbf{q} = \frac{1}{\mu} \mathbf{k} \cdot \mathbf{h} \quad \rightarrow \quad  q_i \mathbf{e}_i = \frac{1}{\mu} \mathbf{k}_{ij} (\mathbf{e}_i \otimes \mathbf{e}_j) \cdot h_k \mathbf{e}_k = \frac{1}{\mu} \mathbf{k}_{ik} h_k \mathbf{e}_i 
\end{equation}
where $h_i = - \frac{\partial p}{\partial x_i} = - p_{,i}$ is the hydraulic gradient. Three numerical experiments must be performed to determine the nine entries in the coefficient matrix of the second order permeability tensor. Thereby, the following pressure gradients are applied for the different cases 
\begin{equation}\label{eq:presssure_bc_appendix}
    a) \,\, h_{1} \neq 0\, \wedge \, h_{2} = h_{3} = 0\, ; \quad 
    b) \,\, h_{2} \neq 0\, \wedge \, h_{1} = h_{3} = 0\, ; \quad
    c) \,\, h_{3} \neq 0\, \wedge \, h_{1} = h_{2} = 0\, ; 
\end{equation}
and we measure three fluxes $q_i$ for each case. By comparing the coefficients for the three equations for the flow $q_i$ (Eq.~(\ref{eq:darcy_appendix}), right) we receive
\begin{equation}
    \begin{aligned}
        q_i  = - \frac{1}{\mu} k_{ik} h_{k} \qquad \rightarrow \qquad & i) & q_1 & = & -\frac{1}{\mu} \left[ k_{11} h_{1} + k_{12} h_{2} + k_{13} h_{3} \right] \\
        \rightarrow \qquad & ii) & q_2 & = & -\frac{1}{\mu} \left[ k_{21} h_{1} + k_{22} h_{2} + k_{23} h_{3} \right] \\
        \rightarrow \qquad & iii) & q_3 & = & -\frac{1}{\mu} \left[ k_{31} h_{1} + k_{32} h_{2} + k_{33} h_{3} \right]
    \end{aligned}
\end{equation}
wherefrom arise the following nine equations by means of the constraints \textit{a)} - \textit{c)} given in Eq.~(\ref{eq:presssure_bc_appendix}):

\begin{equation}\label{eq:perm_computation_all_cases}
    \begin{aligned}
    a) \quad & q_1 = -\frac{1}{\mu} k_{11} h_{1} \, , \quad & \quad q_2 = -\frac{1}{\mu} k_{21} h_{1} \, , \quad & \quad  q_3 = -\frac{1}{\mu} k_{31} h_{1} \, , \\
    b) \quad & q_1 = -\frac{1}{\mu} k_{12} h_{2} \, , \quad & \quad q_2 = -\frac{1}{\mu} k_{22} h_{2} \, , \quad & \quad  q_3 = -\frac{1}{\mu} k_{32} h_{2} \, , \\
    c) \quad & q_1 = -\frac{1}{\mu} k_{13} h_{3} \, , \quad & \quad q_2 = -\frac{1}{\mu} k_{23} h_{3} \, , \quad & \quad  q_3 = -\frac{1}{\mu} k_{33} h_{3} \, . 
    \end{aligned}
\end{equation}
\noindent It results in nine equations for nine unknown entries of the coeffiecient matrix of the second order permeability tensor. Numerically determined, one obtains slightly different values for two corresponding secondary diagonal elements. These are therefore averaged $ k_{12} = k_{21} = \frac{1}{2} (k_{12} + k_{21}) $. The same procedure is adopted for $k_{13}$, $k_{31}$ and $k_{23}$, $k_{32}$.

\bibliographystyle{apacite}
\bibliography{./literature.bib}

\end{document}